\documentclass[11pt,a4paper]{article}
\usepackage{jcappub}

\usepackage{microtype}
\usepackage{amstext,amsthm,amsfonts,slashed}
\usepackage{mathtools}
\usepackage{physics}
\usepackage{braket}
\usepackage{simplewick}

\usepackage[version=4]{mhchem}
\usepackage{siunitx}

\usepackage{caption}
\usepackage{subcaption}
\usepackage{booktabs}
\usepackage{dcolumn}
\usepackage{array}
\usepackage{float}

\usepackage{bm}
\usepackage{bbm}
\usepackage{dsfont}

\usepackage{tikz}
\usepackage[compat=1.1.0]{tikz-feynman}

\usepackage[usenames,dvipsnames]{xcolor}

\usepackage{csquotes}
\usepackage{comment}

\usepackage[nameinlink]{cleveref}

\crefname{equation}{Eq.}{Eqs.}
\Crefname{equation}{Eq.}{Eqs.}
\crefname{appendix}{Appendix}{Appendices}
\Crefname{appendix}{Appendix}{Appendices}

\hypersetup{colorlinks=true,linkcolor=blue!75!black,citecolor=blue!75!black,urlcolor=blue!75!black}

\newcommand{\supsetsim}{\mathrel{\ooalign{\raise.4ex\hbox{$\supset$}\cr\raise-.9ex\hbox{$\sim$}}}}

\newcolumntype{C}[1]{>{\centering\arraybackslash}p{#1}}

\newcommand{\orcid}[1]{\href{https://orcid.org/#1}{\raisebox{+0.42\height}{\includegraphics[height=1.25ex,width=1.25ex]{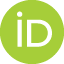}}}}

\author[a]{Juan P.\ Garc\'es\,\orcid{0000-0002-6933-8750},\,}
\author[a,b]{Jisuke Kubo\,\orcid{0000-0003-2211-4685},\,}
\author[a]{Manfred Lindner\,\orcid{0000-0002-3704-6016},\,}
\author[a]{and Markus Reinig\,\orcid{0009-0004-7339-2057}\,}
\affiliation[a]{Max-Planck-Institut f\"ur Kernphysik,\\
P.O. Box 103980, D-69029 Heidelberg, Germany}
\affiliation[b]{Department of Physics, University of Toyama,\\
3190 Gofuku, Toyama 930-8555, Japan}
\emailAdd{juan.garces@mpi-hd.mpg.de}
\emailAdd{kubo@mpi-hd.mpg.de}
\emailAdd{lindner@mpi-hd.mpg.de}
\emailAdd{markus.reinig@mpi-hd.mpg.de}
\title{Dark matter in scale-invariant gravity with hidden-sector condensation}
\abstract{
The origin of the electroweak scale, cosmic inflation, and dark matter are often treated as independent problems beyond the Standard Models of particle physics and cosmology. In this work, we explore the possibility that they instead arise from a common underlying framework based on classically scale-invariant quadratic gravity coupled to a strongly interacting hidden sector. The $R^2$ term naturally realizes Starobinsky inflation, while confinement in the hidden sector dynamically generates the Planck scale and triggers electroweak symmetry breaking through a gravitationally induced Higgs mass generation mechanism. The scalar degree of freedom associated with the $R^2$ term subsequently reheats both the visible and hidden sectors through universal couplings to the energy-momentum tensor, leading to the gravitational freeze-in production of hidden-sector states. We investigate three representative realizations of the hidden sector in which the dark matter candidate is either a hidden $\eta'$ meson, a hidden vector boson, or charged hidden pions, and derive the corresponding dark matter and dark radiation relic abundances.
}
\begin{document}

\maketitle


\newpage

\section{Introduction}

Despite their remarkable experimental success, the Standard Model of particle physics (SM) and the Lambda cold dark matter ($\Lambda$CDM) model of cosmology leave several fundamental questions unanswered. Among them are the origin of the electroweak scale, the nature of dark matter, gravity, and the mechanism responsible for the inflationary expansion of the early Universe. Although these problems are usually addressed independently, it is natural to ask whether they could instead originate from a common underlying principle.

These issues share in some sense a common root: the origin of scales. To explain their emergence and calculability, explicit scales must be avoided in the defining Lagrangian of a Quantum Field Theory (QFT) in four dimensions. This leads to the particularly appealing possibility that the fundamental theory possesses classical scale invariance. In such a framework, no dimensionful parameters appear at the classical level, and all physical mass scales must be generated dynamically~\cite{Coleman:1973jx,Gildener:1976ih} (see also~\cite{Hempfling:1996ht,Meissner:2006zh,Foot:2007as,Chang:2007ki,Foot:2007ay,Foot:2007iy,Iso:2009ss,Alexander-Nunneley:2010tyr,Holthausen:2013ota,Englert:2013gz,Kubo:2014ova,Kubo:2014ida,Lindner:2014oea,Kubo:2015joa,Hatanaka:2016rek,Ahmed:2025gww}). This idea is supported by the observation that the only explicit mass parameter in the SM is the Higgs mass term, whose smallness compared to the Planck scale constitutes the essence of the electroweak hierarchy problem~\cite{Bardeen:1995kv}. If classical scale invariance is a fundamental symmetry of nature, neither the electroweak scale nor any other mass scale can be introduced by hand. Instead, they should emerge through quantum effects, similar to dimensional transmutation in quantum chromodynamics (QCD).

Quadratic gravity~\cite{Stelle:1976gc,Fradkin:1981iu,Salvio:2018crh} provides a particularly attractive realization of this idea~\cite{Salvio:2014soa,Einhorn:2014gfa}. The gravitational action constructed from the curvature-squared operators contains only dimensionless couplings and is therefore classically scale invariant. Moreover, the $R^2$ term introduces an additional scalar degree of freedom, the scalaron, whose dynamics naturally produces Starobinsky inflation~\cite{Whitt:1984pd,Maeda:1987xf,Starobinsky:1980te}. Inflation therefore arises as a direct consequence of the gravitational sector rather than requiring the introduction of an ad hoc inflaton field (see e.g. Ref.~\cite{Aoki:2021skm} and references therein).

In order to realize the dynamical generation of scales by quantum effects, we employ a new strongly interacting hidden gauge sector, analogous to QCD, whose fermionic and/or gluon condensates generate new dynamical scales~\cite{Hur:2011sv,Holthausen:2013ota,Kubo:2014ova,Hatanaka:2016rek,Kubo:2018vdw}. This will generate the largest scale in the theory, the Planck scale. Under the assumption that the gluon condensate dominates over the fermionic one, the Higgs mass parameter is induced through a non-minimal coupling to gravity, providing a dynamical origin for electroweak symmetry breaking. In this picture, the electroweak scale is not fundamental but instead emerges from hidden strong dynamics communicating with the SM through gravity~\cite{deboer2025gravityhierarchyproblem}.

The existence of stable particles is a natural consequence of introducing such a hidden confining sector. Just as QCD predicts pions, vector mesons, and baryons, a hidden QCD-like confining theory naturally gives rise to composite states\footnote{We emphasize that the dark matter candidates of interest need not necessarily be composite states, and a realization where the fundamental gauge boson of the hidden sector acts as dark matter is explored below.} whose stability may follow from exact gauge symmetries, accidental global symmetries, or anomaly structures~\cite{Kribs:2016cew}. This naturally leads to dark matter without the need to impose further stabilizing symmetries~\cite{Nussinov:1985xr}, but instead it appears as a generic prediction of the hidden strongly interacting sector responsible for generating the electroweak scale~\cite{Kubo:2014ida}. Similarly, such a framework also generally leads to a dark radiation abundance~\cite{Ackerman:2008kmp}. Depending on the symmetry structure of the hidden sector, massless gauge bosons or Nambu-Goldstone bosons (NGBs) may survive after confinement and contribute to the effective number of relativistic degrees of freedom~\cite{Weinberg:2013kea}, establishing a direct connection between collider-independent cosmological observables and the dynamics of the hidden sector.

We will show that the scalaron plays a central role in connecting all these phenomena. After inflation, it oscillates around the minimum of its potential and reheats the Universe through its universal coupling to the trace of the energy-momentum tensor~\cite{Gorbunov:2010bn}. Consequently, both the SM and the hidden sector are populated through scalaron decays. Since the hidden sector communicates only gravitationally with the visible sector, the relic abundance of dark matter is naturally produced through gravitational freeze-in~\cite{Garny:2015sjg,Tang:2016vch,Aoki:2021skm}. The same reheating process determines the amount of dark radiation generated in the hidden sector. Inflation, reheating, dark matter production, and dark radiation are therefore not independent ingredients but rather different manifestations of the same underlying dynamics.

In this work, we study the generation of dark relic abundances within different realizations of the general idea described above, i.e. based on classically scale-invariant quadratic gravity coupled to a strongly interacting hidden sector. After describing the general aspects of the framework in~\Cref{sec: the framework} and the onset of the inflationary dynamics in~\Cref{sec: inflation}, we derive general expressions for the gravitational freeze-in of hidden-sector particles produced by scalaron decays in~\Cref{sec: dark relic abundances}. We then explore three representative realizations of the hidden sector in~\Cref{sec: DM models}. In the first model, the dark matter candidate is the hidden $\eta'$ meson arising from a single-flavor confining theory. The second model considers massive hidden gauge vector bosons from an additional hidden local $SU(2)$ symmetry. The third model contains charged hidden pions stabilized by an unbroken hidden $U(1)$ gauge symmetry together with a dark radiation component. For each scenario, we determine the parameter space compatible with the observed dark matter abundance and discuss the associated predictions for dark radiation. Finally, in~\Cref{sec: Summary and conclusions} we summarize the main findings of the present analysis.
%
%
\section{The Framework}\label{sec: the framework}
\begin{figure}[t!]
    \centering
    \includegraphics[width=0.8\linewidth]{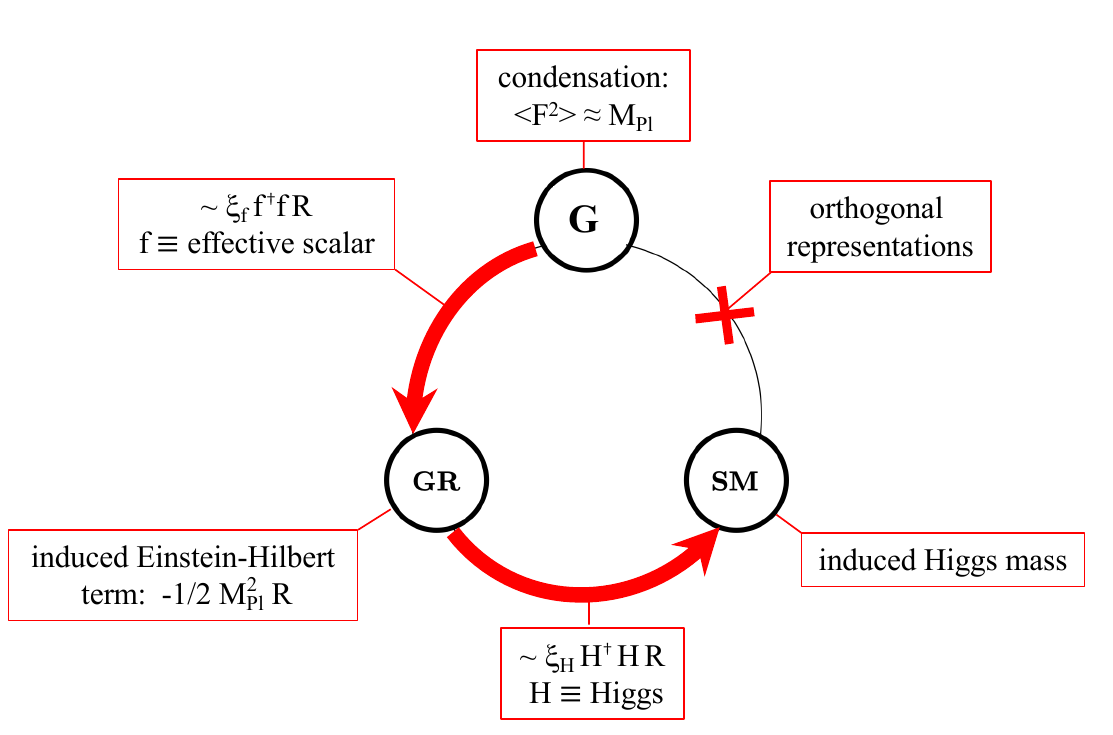}
    \caption{Schematic visualization of the scale generation sequence. First, the Planck scale $M_{\text{Pl}}$ is spontaneously generated by hidden strong dynamics via dimensional transmutation in the $G$-sector. This breaks scale invariance and induces the Einstein-Hilbert term in the quadratic gravity (GR) sector. Since the hidden sector $G$ and the SM sector are completely orthogonal (red cross), the breaking of scale symmetry is communicated to the SM sector only via gravitational interactions. Consequently, the induced electroweak scale in the SM sector is small compared to the Planck scale, realizing a naturally small Higgs mass.}
    \label{fig:3:mechanism}
\end{figure}

In this section, we describe the model of interest in detail, in particular how finite scales arise, and how the appearance of the hierarchy problem can be avoided by a two-step process of scale generation, which is summarized in~\Cref{fig:3:mechanism}. Our model consists of three distinct sectors:
\begin{itemize}
    \itemsep0pt
    \item The SM in the scale-invariant limit, i.e. with vanishing Higgs mass parameter $\mu_{H}$. 
    \item A new scale-invariant hidden sector $G$ with non-abelian gauge interactions 
    \item Scale-invariant quadratic gravity (QG)
\end{itemize}
For the strongly interacting hidden $G$-sector, we use a non-abelian gauge group $SU(N_{c})$, where $N_{c}$ denotes the number of ``hidden colors'' in analogy to QCD. The joint Lagrangian of the hidden and SM sectors is given by
\begin{equation}\label{Lmatter}
    \frac{{\cal L}_\text{matter}}{\sqrt{-g}} = \frac{{\cal L}_\text{SMGF}}{\sqrt{-g}} + D_\mu H^\dag D^\mu H -\lambda_H(H^\dag H)^2 -\frac{1}{2}\Tr F^2 \,,
\end{equation}
where ${\cal L}_\text{SMGF}$ denotes all terms containing only SM gauge and fermionic fields and $F$ is the field-strength tensor of the non-abelian hidden gauge group. Hidden-sector chiral fermions $\psi_{i}$ are not necessary to describe the observed early Universe dynamics, but can, however, be included to naturally generate viable dark matter candidates, as described in~\Cref{sec: Model II}. Moreover, to avoid the hierarchy problem, it is important that the hidden sector is chosen to be completely orthogonal to the scale-invariant SM. This implies that all fermions of the model are either singlets under the SM or under $G$. It also requires that no fundamental scalar field that would allow a Higgs portal term is to be admitted in the hidden sector.

The third sector corresponds to scale-invariant quadratic gravity (QG)~(see~e.g.~\cite{Salvio:2014soa}), which is perturbatively renormalizable~\cite{Stelle:1976gc} and whose Lagrangian in the Jordan frame (indicated by the subscript J) is given by\footnote{As a total covariant derivative, the Gauss-Bonnet term can safely be neglected in our model. Furthermore, terms proportional to the Riemann tensor squared $R_{\mu \nu \rho \sigma}R^{\mu \nu \rho \sigma}$ and the Ricci tensor squared $R_{\mu \nu} R^{\mu \nu}$ can be rewritten as linear combinations of the squared Ricci scalar, the squared Weyl tensor and the Gauss-Bonnet term and can therefore be omitted as well.}\textsuperscript{,}\footnote{The vierbein formalism required by the presence of fermions in the theory is left implicit in our notation.}
\begin{equation}\label{eq: qc}
    \frac{\mathcal{L}_{\text{QG},J}}{\sqrt{-g_{J}}} = \gamma R^{2}_{J} - \kappa C_{\mu \nu \rho \sigma} C^{\mu \nu \rho \sigma}\,,
\end{equation}
where $R$ denotes the Ricci curvature scalar, $C_{\mu \nu \rho \gamma}$ the Weyl tensor, and $\gamma$ and $\kappa$ are dimensionless coupling constants. To maintain renormalizability, the non-minimal coupling to the Ricci scalar
\begin{equation}
    \mathcal{L} \supset -\sqrt{-g_{J}}\, \xi_{H} H^{\dagger} H R_{J}\,,
\end{equation}  
where $\xi_{H}$ denotes the dimensionless non-minimal coupling, must also be included \cite{Stelle:1976gc}. Note that this term implies an unavoidable portal between the SM and gravity, which will become important later on. 

We emphasize that the model respects classical scale invariance, which is motivated by the observation that a theory containing mass terms in its fundamental Lagrangian is unable to explain their origin. The Lagrangian has therefore only dimensionless parameters and all mass scales, including the electroweak scale and the Planck scale, whose vast separation constitutes the gauge hierarchy problem (see e.g. \cite{Craig:2022eqo,Peskin:2025lsg,Garces:2025rgn,Wells:2025hur}), must be generated spontaneously via dimensional transmutation~\cite{Coleman:1973jx}. The resulting breaking sequence of our model is described in the following subsections, and an overview is provided in \autoref{fig:3:mechanism}.

Despite scale invariance being hardly broken by the scale anomaly \cite{Callan:1970yg, Symanzik:1970rt}, it was proven that massless theories exist in perturbation theory \cite{Lowenstein:1975rf, Lowenstein:1975rg}. Consequently, in addition to the scale anomaly, scale invariance must be broken spontaneously to generate physical masses (\cite{Callan:1970ze, Chanowitz:1972da, Chanowitz:1972vd}). Finally, further motivation for scale invariance can be found in cosmology, with the latest constraint for the tensor-to-scalar ratio $r < 0.0032$ \cite{Tristram:2021tvh} indicating that the inflationary potential must be very flat, which is realized in (approximate) scale invariant models, e.g. the  Starobinsky \cite{Starobinsky:1980te} and Higgs inflation models (see e.g. \cite{Rubio:2018ogq} and references therein).
\subsection{Generation of the Planck Mass via Dimensional Transmutation}
The Planck mass can be generated in the non-abelian strongly interacting hidden sector $G$ via dimensional transmutation, similar to QCD in the SM. A gluon condensate $\langle F^{2}\rangle$, where $F_{\mu \nu}^{a}$ denotes the field strength tensor of the hidden sector gauge group, forms at energy scale $\Lambda_{F}$ when the running gauge coupling becomes non-perturbative, resulting in the spontaneous breaking of scale invariance. In a curved spacetime, the Planck mass $M_{\text{Pl}}$ (and hence the Einstein-Hilbert term) and the cosmological constant\footnote{In this work, we do not concern ourselves with the cosmological constant $\Lambda$ and assume that a solution to the cosmological constant problem compatible with our framework exists.} $\Lambda$, can be induced by the dynamical breaking of scale symmetry (see e.g. Ref.~\cite{1982_Adler} for a review), which is supported by lattice computations in the case of a pure Yang-Mills theory~\cite{Donoghue_2018}. 

If chiral fermions are included in the hidden sector, scale invariance is broken by both the hidden gluon condensate $\langle F^{2} \rangle$ and the chiral condensate $\langle \bar{\psi}_{i} \psi_{i}\rangle$, while chiral symmetry is only broken by the latter. However, it was found that the contribution of the chiral condensate induces an Einstein-Hilbert term with the opposite sign to the one required for a positive Newton constant (see e.g. Refs.~\cite{1992_Hill,Inagaki:1993ya, Inagaki:1997kz}). Therefore, if a hidden sector with chiral fermions is included, one must assume that the gluon contribution dominates. Further motivation for this assumption comes from the observation that the hidden gauge sector is responsible for generating the dynamical scale through dimensional transmutation. In a classically scale-invariant gauge theory, the spontaneous breaking of scale invariance is associated with the scale anomaly and is therefore controlled by the $\beta$-function of the hidden gauge coupling. Analogous to QCD, the gluonic contribution to the $\beta$-function must dominate over the fermionic contribution, which enters with the opposite sign, in order for the $\beta$-function to remain negative and the theory to be asymptotically free. Since the dynamical scale ultimately originates from the gauge dynamics, it is natural to regard the gluon condensate as the primary order parameter associated with scale generation and to assume that its contribution to the induced Planck mass dominates over that of the chiral condensate.

We emphasize that the condensation in the hidden sector $G$ does not spoil its orthogonality with the SM sector, in the sense that no direct interaction is generated as a by-product of confinement. Even though the condensates can be described by effective scalar degrees of freedom in the effective theory, in the absence of gravity no effective Higgs portal coupling must be included, as it can not emerge from the fundamental theory. 
\subsection{Induced Higgs Mass}
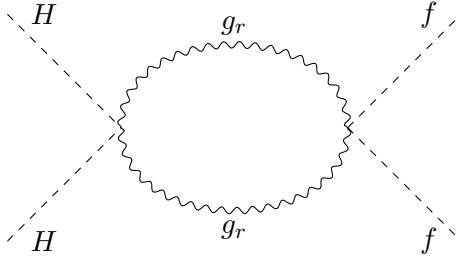
\begin{figure}[t!]
\centering
\begin{tikzpicture}[baseline=(v1.base)]
\begin{feynman}
    \vertex (a1) [at={(-3, 1.5)}, label={[label distance=5pt]right:$H$}];
    \vertex (a2) [at={(-3,-1.5)}, label={[label distance=5pt]right:$H$}];
    \vertex (v1) at (-1.5, 0);
    \vertex (v2) at ( 1.5, 0); 
    \vertex (b1) [at={( 3, 1.5)}, label={[label distance=5pt]left:$f$}];
    \vertex (b2) [at={( 3,-1.5)}, label={[label distance=5pt]left:$f$}];

    \diagram*
    { (a1) -- [scalar] (v1),
      (a2) -- [scalar] (v1),
      (v1) -- [photon, half left, looseness=1.25, edge label=$g_{r}$] (v2),
      (v2) -- [photon, half left, looseness=1.25, edge label=$g_{r}$] (v1),
      (v2) -- [dashed] (b1),
      (v2) -- [dashed] (b2)};
\end{feynman}
\end{tikzpicture} 
\caption{Diagrammatically, the emergence of the induced Higgs mass can be understood by formally introducing a non-minimally coupled, non-propagating \enquote{Froggatt-Nielsen} field $f$. The induced one-loop correction to the Higgs mass parameter is then mediated by the gravitational interaction, i.e. the scalaron and spin-two ghost contained in $g_{r}$. The Higgs mass parameter can then be readily obtained by reinstating the Planck Mass $M_{\text{Pl}}^{2}$ for the non-minimal coupling of the Froggatt-Nielsen field $\xi_{f}f^{2}$. }
\label{fig:4:portal}
\end{figure}

As mentioned above, the bare mass parameter $\mu_{H}$ of the SM Higgs field must vanish due to classical scale symmetry. Moreover, as depicted in~\Cref{fig:3:mechanism}, the SM sector, being completely orthogonal to the confining hidden sector, only feels the breaking of scale invariance through gravitational interactions. To compute the corresponding induced Higgs mass, we formally introduce a non-propagating ``Froggatt-Nielsen'' field $f$ and rewrite the Einstein-Hilbert term according to
\begin{align}
    -\frac{1}{2} M_{\text{Pl}}^{2} R_{J}\longrightarrow-\frac{1}{2} \xi_{f} f^{2} R_J\,.
\end{align}
With this notation, a Higgs mass is induced by the effective portal coupling
\begin{equation}
    \mathcal{L}_{\text{portal}} = -\sqrt{-g}\,\frac{\lambda_{fH}^{\text{ind}}}{2}\,f^{2} H^{\dagger}H\,,
\end{equation}
and we can use the results found in Ref.~\cite{Salvio2014} for the one-loop $\beta$-function of the induced portal coupling of two non-minimally coupled scalar fields,
\begin{equation}
    \frac{\mathrm{d} \lambda_{fH}^{\text{ind}}}{\mathrm{d} \ln \mu^{2}} = \frac{1}{128\pi^{2}}\,\xi_{f}\,\xi_{H}\left(\frac{5}{\kappa^{2}} + \frac{1}{9 \gamma^{2}} \left(6\xi_{f}+1\right)\left(6\xi_{H}+1\right) \right).
\end{equation} 
The expression is given in terms of the dimensionless parameters $\kappa, \gamma$, and the non-minimal coupling $\xi_{H}$. Reinstating $M_{Pl}^{2}$ in place of $\xi_{f}f^{2}$ yields the induced mass term
\begin{equation}
    \mathcal{L}_{\text{ind.-mass}} = -\sqrt{-g}\, \mu_{H}^{2} H^{\dagger}H,
\end{equation} 
with induced Higgs mass parameter
\begin{equation}\label{eq:mus2}
    - \mu_{H}^{2} = - \frac{\xi_{H}M_{\text{Pl}}^{2}}{256\pi^{2}} \left(\frac{5}{\kappa^{2}} + \frac{1}{9 \gamma^{2}}\, (6\xi_{f}+1)(6\xi_{H}+1) \right) \times \mathcal{C},
\end{equation}
where all non-perturbative corrections are consolidated in the parameter $\mathcal{C}$, assumed to be in the range $\mathcal{O}(10^{-1})$ to $\mathcal{O}(10)$. In \autoref{fig:4:portal}, the induced Higgs mass is diagrammatically visualised at one loop.

Assuming that cosmic inflation in the early Universe is dominated by Starobinsky inflation, requires the parameter $\gamma$ to take values in the range $\gamma \simeq \mathcal{O}(10^8)$--$\mathcal{O}(10^9)$~\cite{Ema:2017rqn, Pi:2017gih, Salvio:2017xul, Gundhi:2018wyz, Kubo:2018kho, Enckell:2018uic, Kubo:2020fdd, Aoki:2021skm, Cecchini:2024xoq}. For $\mathcal{C} \simeq \mathcal{O}(1)$ and $\xi_{H} \simeq \mathcal{O}(1)$, this would yield $\mu_{H}^{2} \simeq O(10^{7})\, \si{\giga\eV}$, significantly larger than the physical value. However, in the semi-conformal limit,\footnote{Not to be confused with the quasi-conformal limit, discussed e.g. in \cite{Salvio:2014soa, Salvio:2019wcp, Salvio:2020axm, Kannike:2015apa}.} where the non-minimal Higgs coupling approaches $\xi_{H} = -1/6$~\cite{Salvio:2014soa, Salvio:2019wcp, Salvio:2020axm, Kannike:2015apa}, the contribution proportional to $\gamma^{-1}$ becomes negligibly small and the observed Higgs mass value is obtained for $\kappa \simeq 5 \times 10^{14}$. If $\mu_{0}$ denotes the energy scale at which $\xi_{H} = -1/6$, corrections to the Higgs mass from the running of the non-minimal coupling are negligible in the energy range $\mathcal{O}(10^{-10}) \lesssim \mu/\mu_{0} \lesssim \mathcal{O}(10^{10})$ (see e.g. Ref.~\cite{deboer2025gravityhierarchyproblem}). Furthermore, it can be demonstrated that the correction obtained from kinetic scalaron-Higgs mixing is of $\mathcal{O}(v_{H}^{6}/M_{\text{Pl}}^{4})$ and is therefore negligible (see e.g. Ref.~\cite{deboer2025gravityhierarchyproblem}). 

To sum up, in contrast to earlier models, the framework described above contains no initial portal coupling that would need to be required to take values of $\mathcal{O}(10^{-32})$ in order to stabilize the hierarchy between the electroweak and Planck scales.
%
%
\section{Inflationary dynamics}\label{sec: inflation}
Within the framework described in~\Cref {sec: the framework}, there are different scalar degrees of freedom that can contribute to the inflationary dynamics. The scalaron field, denoted $\chi$ and $\phi$ in the Jordan and Einstein frames, respectively, is contained in the $R^2$ term of the gravity sector. The SM Higgs doublet $H$ and a composite scalar $\sigma$, a chiral condensate, are present in the visible and hidden matter sectors, respectively. However, as discussed in the previous section, explaining the smallness of the Higgs mass induced from gravitational interactions (see~\Cref{eq:mus2}) requires sitting close enough to the semi-conformal regime, namely $\xi_H\rightarrow-1/6$. In this limit, the Higgs field decouples from the inflationary dynamics and can be safely ignored. As for the composite scalar, resulting from the confining hidden dynamics, the associated effective potential, $V_\text{NJL}(\sigma, \phi)$, can be calculated using the Nambu-Jona-Lasinio model (NJL)~\cite{Nambu:1961fr,Nambu:1961tp}.

In this section, we describe how the associated effective potential containing the scalaron and the NJL condensate can be reduced to single-field Starobinsky inflation by following the analysis of Ref.~\cite{Aoki:2021skm}. We then use the resulting potential to relate parameters and observables relevant for the corresponding inflationary and reheating dynamics.
\subsection{From multi-field to single-field Starobinsky inflation} 
In the self-consistent mean-field approximation, the hidden $SU(N_c)$ strongly interacting sector with $N_f$ fermions in the fundamental representation may be described by the NJL effective potential (see Ref.~\cite{deboer2025gravityhierarchyproblem} for more details)
\begin{align}\label{eq: VNJL sigma}
    V_{\rm NJL}(\sigma) = \frac{N_f}{8G}\sigma^2+N_cN_f\,I_0(\sigma,\Lambda_H)\,,
\end{align}
where $M(\sigma)=\sigma$ is the constituent hidden-fermion mass, $\sigma$ denotes the composite scalar mean field associated with the chiral condensate $\langle\bar\Psi\Psi\rangle$, and
\begin{align}\label{eq: I0-sigma}
    I_0(M(\sigma),\Lambda_H)=\frac{1}{16\pi^2}\left[\sigma^4\ln\!\left(1+\frac{\Lambda_H^2}{\sigma^2}\right)-\Lambda_H^4\ln\!\left(1+\frac{\sigma^2}{\Lambda_H^2}\right)-\Lambda_H^2\sigma^2\right]\,.
\end{align}
The parameter $G$ denotes the four-fermion NJL coupling and $\Lambda_H$ is the ultraviolet cutoff of the NJL description.\footnote{For simplicity, we neglect the multi-fermion interaction responsible for the explicit breaking of the axial $U(1)_A$ symmetry in the underlying hidden QCD-like theory. Such a Kobayashi--Maskawa--'t~Hooft determinant interaction depends on the number of flavors and corresponds to a $2N_f$-fermion operator, reducing to the familiar six-fermion interaction only for $N_f=3$. Since our analysis focuses on the formation of the chiral condensate and the resulting generation of the Planck scale, rather than on the pseudoscalar spectrum, it is sufficient to retain only the four-fermion NJL interaction. This allows us to formulate the effective theory for arbitrary $N_f$ without affecting the subsequent inflationary dynamics.} We define the shifted potential
\begin{align}\label{eq: shifted NJL}
    U(\sigma)=V_{\rm NJL}(\sigma)-V_{\rm NJL}(v_\sigma)\,,
\end{align}
such that the potential energy of the zero-curvature vacuum is zero, and where $v_\sigma=\langle\sigma\rangle$.

Neglecting the Weyl-tensor-squared operator for the moment, the relevant Jordan-frame action can be written schematically as
\begin{align}\label{eq: Jordan NJL scalaron}
    S_J =\int d^4x\,\sqrt{-g_J} \left[-\frac{M_{\rm Pl}^2}{2}R_J+\gamma R_J^2+\frac{Z_\sigma^{-1}}{2}g_J^{\mu\nu}\partial_\mu\sigma\partial_\nu\sigma-U(\sigma)\right]\,,
\end{align}
with $Z_\sigma$ the wave-function normalization of the scalar condensate, and since its detailed form is not relevant for the location of the inflationary valley, we leave it unspecified. The $R_J^2$ term can be linearized by introducing an auxiliary field $\chi$, namely
\begin{align}\label{eq: auxiliary replacement}
    \gamma R_J^2\,\longrightarrow\,\gamma \frac{M_{\mathrm{Pl}}}{\sqrt{6}}\chi R_J-\gamma\frac{M_{\mathrm{Pl}}^2}{24}\chi^2\,.
\end{align}
The equivalence follows directly from the algebraic equation of motion of
$\chi$,
\begin{align}\label{eq: chi eom}
    \frac{\delta S_J}{\delta\chi}=M_{\mathrm{Pl}}\gamma\left(\frac{R_J}{\sqrt{6}}-\frac{M_{\mathrm{Pl}}}{12}\chi\right)=0\quad\Rightarrow\quad\chi=\frac{12}{\sqrt{6}}\frac{R_J}{M_{\mathrm{Pl}}}\,,
\end{align}
which when substituted back into~\Cref{eq: auxiliary replacement} reproduces the original term. The pure-gravity part of the Jordan-frame action may therefore be written as
\begin{align}\label{eq: Jordan action auxiliary}
    S_J \supset\int d^4x\,\sqrt{-g_J}\bigg[-\frac{M_{\rm Pl}^2}{2}\left(1-\frac{2\gamma\chi}{\sqrt{6}M_{\rm Pl}}\right)R_J-\gamma\frac{M_{\mathrm{Pl}}^2}{24}\chi^2\bigg]\,.
\end{align}
Although the auxiliary field $\chi$ has no kinetic term in the Jordan frame, it corresponds to the scalar degree of freedom contained in the metric. Indeed, after expanding the metric around flat spacetime, the trace part of the metric fluctuation may be identified with $\chi$, while the traceless transverse part describes the spin-two degrees of freedom. To make this degree of freedom explicit, we expand the metric around Minkowski spacetime,
\begin{align}
    g_{\mu\nu}=\eta_{\mu\nu}+h_{\mu\nu}\,,
\end{align}
and introduce the rescaled field
\begin{align}
    \tilde\chi\equiv\gamma\chi\,.
\end{align}
At quadratic order, the scalar-metric mixing can be diagonalized by the field redefinition
\begin{align}\label{eq: metric expansion 1}
    h_{\mu\nu} = \hat h_{\mu\nu}+\sqrt{\frac23}\,\frac{\tilde\chi}{M_{\rm Pl}}\eta_{\mu\nu}\,,
\end{align}
where $\hat h_{\mu\nu}$ denotes the spin-two part of the metric fluctuation. More explicitly, the terms relevant at quadratic order are
\begin{align}
    \mathcal L_J^{(2)}=-\frac{M_{\rm Pl}^2}{2}\left[\sqrt{-g}\,R\right]^{(2)}+\frac{M_{\rm Pl}}{\sqrt6}\tilde\chi\,R^{(1)}(h)- \frac{M_{\rm Pl}^2}{24\gamma} \tilde\chi^2\,.
\end{align}
Upon substituting~\Cref{eq: metric expansion 1} and using the replacement in~\Cref{eq: auxiliary replacement}, the scalar-metric mixing is diagonalized and the scalar part becomes
\begin{align}
    \mathcal L_J^{(2)} = \mathcal L_{\rm grav}^{(2)}[\hat h]+\frac12(\partial\tilde\chi)^2 - \frac12m_\phi^2\tilde\chi^2 +\cdots\,,
\end{align}
where the ellipsis denotes higher-order interactions, and terms belonging to the remaining gravitational sector and
\begin{equation}
    m_\phi^2=\frac{M_{\rm Pl}^2}{12\gamma}\,.
\end{equation}
To obtain the Einstein-frame action, we return to the Jordan-frame action in~\Cref{eq: Jordan action auxiliary} and perform the Weyl transformation parametrized by the dimensionless conformal factor
\begin{align}\label{eq: Omega definition}
    \Omega^2(\chi)\equiv1-\frac{2\gamma\chi}{\sqrt{6}M_{\rm Pl}}\,.
\end{align}
The Einstein-frame metric is then defined by
\begin{align}\label{eq: Weyl transformation}
    g_{\mu\nu}=\Omega^2 g^J_{\mu\nu}\,.
\end{align}
Under this transformation, the metric and the square root of its determinant transform as
\begin{align}\label{eq: Weyl metric relations}
    g_J^{\mu\nu}=\Omega^2 g^{\mu\nu}\qquad\mathrm{and}\qquad \sqrt{-g_J}=\Omega^{-4}\sqrt{-g}\,,
\end{align}
respectively, while the Ricci scalar transforms as
\begin{align}\label{eq: Ricci Weyl}
    R_J=\Omega^2\left[R+3\Box\ln\Omega^2-\frac{3}{2}g^{\mu\nu}\partial_\mu\ln\Omega^2\partial_\nu\ln\Omega^2\right]\,.
\end{align}
Inserting~\Cref{eq: Weyl metric relations,eq: Ricci Weyl} into~\Cref{eq: Jordan action auxiliary} with the scalar condensate terms reinstated, and disregarding the term proportional to $\Box\ln\Omega^2$, which is a boundary term, leads to the Einstein-frame action 
\begin{align}\label{eq: Einstein action before phi}
    S_E=\int d^4x\,\sqrt{-g}\bigg[&-\frac{M_{\rm Pl}^2}{2}R+\frac{3M_{\rm Pl}^2}{4}g^{\mu\nu}\partial_\mu\ln\Omega^2\partial_\nu\ln\Omega^2\\\nonumber&+\frac{Z_\sigma^{-1}}{2}\Omega^{-2}g^{\mu\nu}\partial_\mu\sigma\partial_\nu\sigma-\Omega^{-4}\left(U(\sigma)+\gamma\frac{M_{\rm Pl}^2}{24}\chi^2\right)\bigg]\,.
\end{align}
Next, we introduce the canonically normalized scalaron in the Einstein frame $\phi$ via the definitions
\begin{align}\label{eq: scalaron definition}
    \phi\equiv\sqrt{\frac{3}{2}}\,M_{\rm Pl}\ln\Omega^2\qquad\mathrm{and}\qquad\Phi(\phi)\equiv\sqrt{\frac{2}{3}}\frac{\phi}{M_{\rm Pl}}\,,
\end{align}
so that $\Omega^2=e^{\Phi(\phi)}$ and
\begin{align}
    \frac{3M_{\rm Pl}^2}{4}\left(\partial\ln\Omega^2\right)^2=\frac{1}{2}(\partial\phi)^2\,.
\end{align}
\Cref{eq: Omega definition} can now be inverted to eliminate the auxiliary field
\begin{align}\label{eq: chi in terms of phi}
    \chi=\frac{\sqrt{6}M_{\rm Pl}}{2\gamma}\left(1-\Omega^2\right)=\frac{\sqrt{6}M_{\rm Pl}^2}{2\gamma}\left(1-e^{\Phi(\phi)}\right)\,,
\end{align}
and the complete Einstein-frame action becomes
\begin{align}\label{eq: Einstein action sigma phi}
    S_E=\int dx^4 \,\sqrt{-g}\bigg[-\frac{M_{\rm Pl}^2}{2}R+\frac{g^{\mu\nu}}{2}\partial_\mu\phi\partial_\nu\phi+\frac{e^{-\Phi(\phi)}}{2}Z_\sigma^{-1}g^{\mu\nu}\partial_\mu\sigma\partial_\nu\sigma-V_E(\sigma,\phi)\bigg]\,,
\end{align}
with
\begin{align}\label{eq: Einstein potential}
    V_E(\sigma,\phi)= e^{-2\Phi(\phi)}U(\sigma)+\frac{M_{\rm Pl}^4}{16\gamma}\left(1-e^{-\Phi(\phi)}\right)^2.
\end{align}
The first term is the Weyl-rescaled NJL potential and the second is the standard Starobinsky potential. Notice also that the kinetic term of the composite scalar is multiplied by $e^{-\Phi}$, so that the two-field system has a non-trivial field-space metric even though the potential takes a relatively simple form.

The bottom of the potential of~\Cref{eq: Einstein potential} in the $\sigma$ can be obtained from solving
\begin{align}\label{eq: sigma derivative}
    \frac{\partial V_E}{\partial\sigma}=e^{-2\Phi(\phi)}\frac{\partial U}{\partial\sigma}=0\,.
\end{align}
Since the Weyl factor is non-zero, the valley condition is equivalent to
\begin{align}\label{eq: sigma valley NJL}
    \left.\frac{\partial U}{\partial\sigma}\right|_{\sigma=\sigma_v}=0\,,
\end{align}
from which it follows that the location of the valley does not depend on the scalaron, i.e. $\sigma_v(\phi)=v_\sigma$.\footnote{We note that the result $\sigma_v(\phi)=v_\sigma$ follows exactly from the truncated two-field potential in~\Cref{eq: Einstein potential} because the coefficients of $R$ and $R^2$ have been taken to be independent of $\sigma$. Non-minimal couplings of the condensate to curvature, curvature-dependent corrections to the NJL action, or a $\sigma$-dependent coefficient of the $R^2$ operator would generally bend the valley and yield a non-trivial trajectory $\sigma_v(\phi)$. Furthermore, the existence of a stationary valley does not by itself guarantee single-field dynamics. The approximation requires the fluctuation orthogonal to the trajectory to be sufficiently heavy,
\begin{align}
    m_{\sigma,\perp}^2(\phi)\gg H^2(\phi)\,,\nonumber
\end{align}
and the trajectory to have a negligible turning rate in field space. Consequently, the reduction to~\Cref{eq: Starobinsky from NJL} should be understood as valid in the parameter region where the hidden condensate possesses a stable and sufficiently steep minimum, rather than as a model-independent consequence of confinement.} Substituting this into the full potential yields
\begin{align}\label{eq: potential along valley}
    V(\phi)&\equiv V_E(v_\sigma,\phi)\nonumber\\&=e^{-2\Phi(\phi)}U(v_\sigma)+\frac{M_{\rm Pl}^4}{16\gamma}\left(1-e^{-\Phi(\phi)}\right)^2\,,
\end{align}
which with the vacuum-energy convention $U(v_\sigma)=0$, reduces to
\begin{align}\label{eq: Starobinsky from NJL}
    V(\phi)=\frac{M_{\rm Pl}^4}{16\gamma}\left(1-e^{-\sqrt{2/3}\,\phi/M_{\rm Pl}}\right)^2\,,
\end{align}
which is the standard potential for Starobinsky inflation~\cite{Starobinsky:1980te,Mukhanov:1981xt,Starobinsky:1983zz,Maeda:1987xf} in terms of the dimensionless parameter of the $\gamma R^{2}$ term.
\begin{figure}[t!]
\begin{center}
\includegraphics[width=0.49\linewidth]{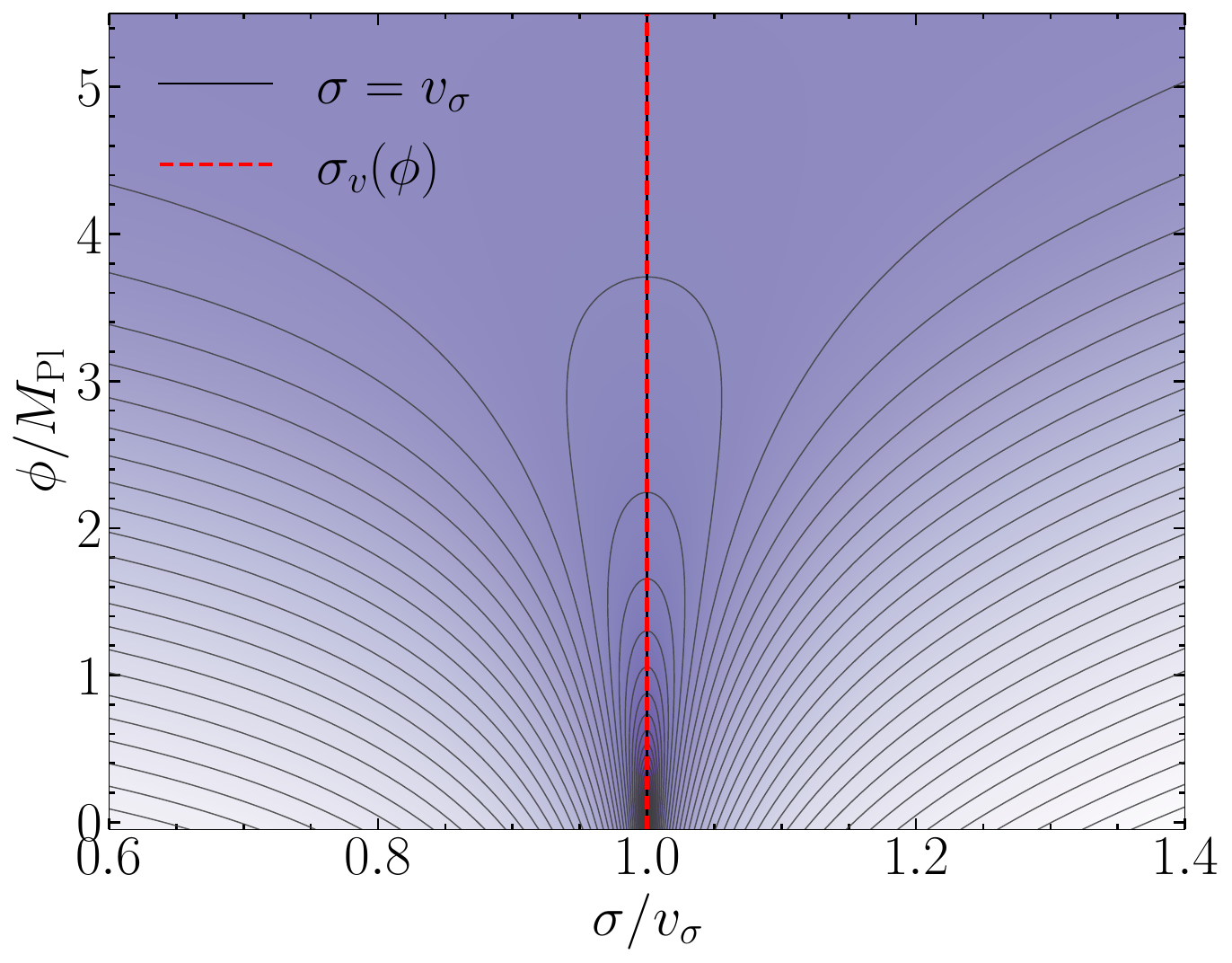}
\includegraphics[width=0.49\linewidth]{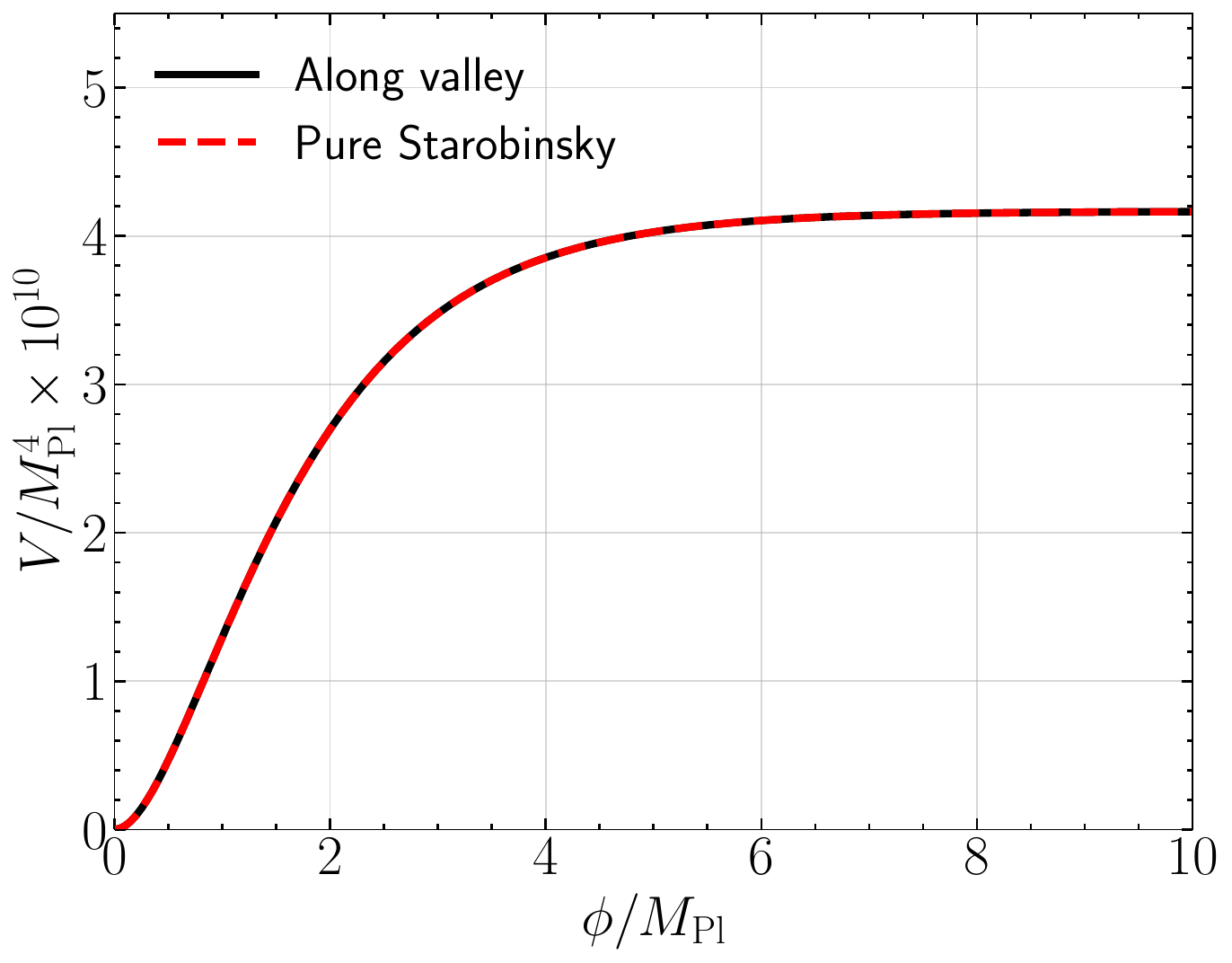}
\caption{\textit{Left:} The effective potential $V_\text{NJL}(\sigma,\phi)$. It is calculated in the NJL model for representative values of the parameters. \textit{Right:} The inflationary potential along the valley of $V_{\rm NJL}(\sigma,\phi)$ where $\sigma(\phi)\simeq v_\sigma$ (black) and the equivalent pure Starobinsky potential with $\sigma=v_\sigma$.}
\label{fig:sigma}
\end{center}
\end{figure}

\Cref{fig:sigma} shows the inflationary potential for a representative set of parameters. The left panel shows that $\sigma$ is indeed frozen along the inflationary trajectory and, as a consequence, the three-field scalar potential relevant for inflation effectively reduces to a single-field system exhibiting Starobinsky inflation, as shown in the right panel.
\subsection{Inflationary parameters}
The spectral index $n_s$ and the tensor-to-scalar ratio $r$ for the inflationary potential in~\Cref{eq: Starobinsky from NJL} are given in terms of the number of e-foldings during inflation $N_e$ by
\begin{align}
    n_s\simeq1- \frac{2}{N_e}\,,\quad \mathrm{and}\quad r\simeq \frac{12}{N_e^2}\,.
    \label{eq: spectral}
\end{align}
The measured value of the scalar amplitude~\cite{Aghanim:2018eyx,Planck:2018jri}\,,
\begin{align}\label{eq: scalar amp meas}
    A_s = e^{3.044\pm0.014}\times 10^{-10}\,,
\end{align}
constrains the value of $\gamma$ through the relation
\begin{align}\label{eq: scalar amp}
    A_s=\frac{V(\phi_*)}{24\pi^2 \varepsilon_* M^4_\text{Pl}}\,,
\end{align}
valid in the slow-roll approximation, with $\phi_*$ the field value at the CMB horizon exit~\cite{Aghanim:2018eyx}, and 
\begin{align}\label{eq: slow-roll param}
    \varepsilon_*\simeq \frac{M^2_\text{Pl}}{2}\left( \frac{V'(\phi^*)}{V(\phi^*)}\right)^2,
\end{align}
the slow-roll parameter evaluated at $\phi_*$. Combining \Cref{eq: scalar amp meas,eq: scalar amp} leads to a range of allowed values of $\gamma$ as a function of the number of e-foldings, namely
\begin{align}
    \gamma\in[4.5,\,6.4] \times 10^8\,,\quad\mathrm{for}\quad N_e\in[49,\,59]\,.
\end{align}
\begin{figure}
  \centering
  \includegraphics[width=0.5\linewidth]{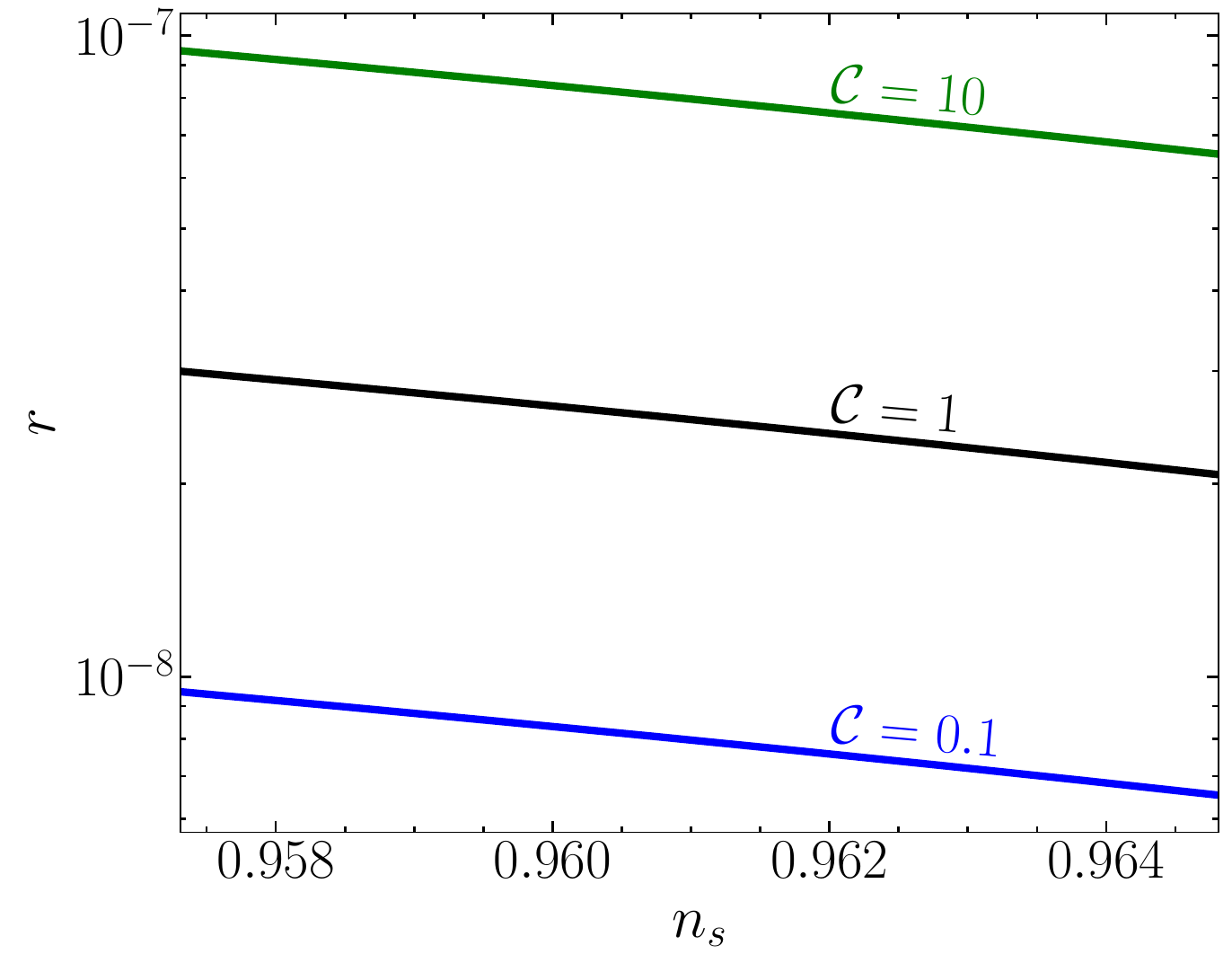}
  \caption{Tensor-to-scalar ratio as a function of the spectral index in the presence of a spin-2 ghost, for different values of the coefficient $\cal C$.}
  \label{r-weyl}
\end{figure}
We note that the above results hold only in the absence of the Weyl-tensor-squared term (\Cref{eq: qc}). When this term is included, the corresponding spin-two ghost modifies the prediction of $r$~\cite{Deruelle:2012xv,Myung:2014jha,Myung:2015vya,Salvio:2017xul,
Ghilencea:2019rqj,Anselmi:2020lpp,Bianchi:2025tyl}, as well as the tensor spectral index $n_t$ according to~\cite{Salvio:2020axm,Anselmi:2020lpp,Kubo:2025jla,Bianchi:2025tyl}
\begin{align}\label{eq: ratio}
    r=16\,\varepsilon_*\to r= \frac{16\,\varepsilon_*}{1+2 {\cal H}^2_*/m^2_\text{gh}}\,,\quad
    n_t = -2\, \varepsilon_* \to n_t=\frac{-2 \,\varepsilon_* }{1+2 {\cal H}^2_*/m^2_\text{gh}}\,,
\end{align}
where ${\cal H}_*$ is the Hubble parameter at horizon exit and $m_\text{gh}=M_\text{Pl}/\sqrt{4\kappa}$ is the mass of the spin-two ghost.\footnote{We ignore the controversial debate regarding whether an undesirable growth of the scalar part of the ghost perturbation in the superhorizon regime is a gauge artifact \cite{Deruelle:2010kf,Ivanov:2016hcm,Salvio:2017xul} or not \cite{DeFelice:2023psw}.} Requiring~\Cref{eq:mus2} to yield the desired induced Higgs mass squared, $m^2_H =-2\mu_H^2\simeq (125\,\mbox{GeV})^2q $, in the limit $\xi_H\to-1/6$ results in
\begin{align}
    \kappa\simeq 5.0\times 10^{14}\, {\cal C}^{1/2}\,.
\end{align}
With this, $r$ can be computed as a function of $n_s$ using~\Cref{eq: ratio}, as shown in~\Cref{r-weyl} for different values of $\cal{C}$. The effect of the spin-two ghost is significant as it reduces $r$ by at least four orders of magnitude.
\subsection{Relating the reheating temperature to the duration of inflation}
The inflationary period described above ends when the scalaron field $\phi$ starts oscillating around its minimum and reheats the Universe. Since $\phi$ couples to both visible and dark particles, it generates visible and dark relic abundances. Depending on the masses of the dark daughter fields, they will contribute to the dark matter or dark radiation abundances, with the corresponding energy densities being a function of the reheating temperature. As shown in~\Cref{eq: spectral}, the inflationary parameters are a function of the number of e-foldings during inflation.  On the other hand, the observables of interest, i.e. the dark matter and dark radiation abundances, depend on the reheating temperatures of the visible and dark sectors. Thus, it becomes useful to find relations between $N_e$ and these reheating temperatures.

In principle, the hidden sector particles could also be produced before or during inflation. However, since their energy densities are diluted by a factor of $(e^{-N_e})^q \simeq10^{-22 q}\sim 10^{-26 q}$, where $q=3 (4)$ for dark matter (dark radiation), they can be ignored at the end of inflation. We therefore assume that any relic abundance generated before reheating is negligible and that the non-negligible component comes solely from the decays of the scalaron. Furthermore, the interactions between the dark and SM sectors are gravitationally suppressed and can be safely ignored. On the other hand, dark matter and dark radiation particles can generally interact with each other, as will be the case in the model studied in~\Cref{sec: model III}. However, in the case where these interactions are also weak, they can be safely ignored.

Since the reheating temperature of dark radiation $T_\text{RH}^\text{DR}$ and that of the SM bath $T_\text{RH}^\text{SM}$, both to be defined below, will generally be different, we introduce the ratio
\begin{align}\label{eq: zeta}
    \zeta =& \frac{T_\text{RH}^\text{DR}}{ T_\text{RH}^\text{SM}}\,.
\end{align}
The energy density at the end of reheating, given by the moment at which the inflaton energy density is comparable to that of the daughter fields, can be related to the energy density at the end of inflation through~\cite{Liddle:2003as,Martin:2010kz,Lozanov:2017hjm,Planck:2018jri}
\begin{align}\label{eq: rhoend rhorh}
    \frac{k_*}{a_0 {\cal H}_0}&=\frac{a_*{\cal H}_*}{a_0 {\cal H}_0}=
    \left( \frac{a_*}{a_\text{end}} \right)\left( \frac{a_\text{end}}{a_\text{RH}} \right)
    \left( \frac{a_\text{RH} \,\rho^{1/4}_\text{RH}}{\sqrt{3}\,a_0\, {\cal H}_0} \right)
    \left( \frac{\rho^{1/4}_\text{end}}{\rho^{1/4}_\text{RH}} \right)
    \left( \frac{\sqrt{3}\,{\cal H}_*}{\rho^{1/4}_\text{end}} \right)\,,
\end{align}
where $k_*=a_* {\cal H}_*$ is the pivot scale, and ${()}_*$ denotes values at horizon exit. The pivot scale is an arbitrary parameter which is usually set to $k_*=0.002~\text{Mpc}^{-1}$ or $0.05~\text{Mpc}^{-1}$~\cite{Aghanim:2018eyx,Planck:2018jri}. Taking the logarithm of both sides of~\Cref{eq: rhoend rhorh} leads to
\begin{align}\label{eq: N}
    \ln\left(\frac{k_*}{a_0 {\cal H}_0}\right)&=-N_e+\ln R_\text{rad}+\ln\left( \frac{a_\text{RH} \,\rho^{1/4}_\text{RH}}{\sqrt{3}\,a_0\, {\cal H}_0} \right)+\frac{1}{4}\ln\left( \frac{9\,{\cal H}_*^4}{\rho_\text{end}} \right)\,,
\end{align}
where $N_e=\ln(a_\text{end}/ a_*)$ and 
\begin{align}\label{eq: lnR}
    \ln R_\text{rad}\equiv\frac{1}{4}\ln \left(\frac{\rho_\text{end}}{\rho_\text{RH}}\right)+\log\left(\frac{a_\text{end}}{a_\text{RH}}\right)\,.
\end{align}
Using the relation $dN={\cal H} dt$ and defining the equation of state $w$ via $d\ln\rho/dN\equiv-3 (1+w)$, allows to write
\begin{align}\label{eq: lnrho}
    \ln \left(\frac{\rho_\text{end}}{\rho_\text{RH}}\right)
    &=3\int_{N_\text{end}}^{N_\text{RH}} d n \big(1+w(n)\big)
    =3\Delta N(1+\bar{w})\,,
\end{align}
where $\Delta N= N_{\rm RH}-N_{\rm end}$ and with $\bar\omega$ the mean equation of state defined as
\begin{align}\label{w}
    \bar{w}\equiv\frac{1}{\Delta N}\int_{N_\text{end}}^{N_\text{RH}} dn \, w(n)\,.
\end{align}
Moreover, since $\Delta N=-\ln (a_\text{end}/a_\text{RH})$,~\Cref{eq: lnR} can be written as
\begin{align}\label{eq: lnRad1}
    \ln R_\text{rad}=\frac{1}{4}\Delta N(-1+3\bar{w})=\frac{1-3\bar{w}}{12(1+\bar{w})}\ln \left(\frac{\rho_\text{RH}}{\rho_\text{end}}\right)\,.
\end{align}
The mean equation of state can be obtained from the behavior of the potential in the reheating phase, namely, during the period of oscillations near the minimum. For a potential in the reheating phase described by $V(\phi)\propto (\phi/M_\text{Pl})^n$ it reads~\cite{Turner:1983he,Shtanov:1994ce}
\begin{align}\label{eq: mean w}
    \bar{w}=\frac{n-2}{n+2}\,.
\end{align}
Near the minimum, the Starobinsky potential in~\Cref{eq: Starobinsky from NJL} can be expanded as
\begin{align}
    V(\phi)=\frac{M^4_\text{Pl}}{24 \gamma}\left[\left(\frac{\phi}{M_\text{Pl}}\right)^2+O\left(\left(\phi/M_\text{Pl}\right)^3\right)\right]\,,
\end{align}
and thus $\bar{w}\simeq 0$. 

The conservation of entropy per comoving volume allows us to write
\begin{align}\label{eq: s consv}
    s_\text{RH}^\text{SM} \,a^3_\text{RH}=&s_0^\text{SM} \,a^3_0\,,
\end{align}
with $s_{\rm RH}$/$a_{\rm RH}$ and $s_0$/$a_0$ the entropy density/scale factor at the end of reheating and today, respectively, which can be written in terms of the corresponding bath temperature as
\begin{align}\label{eq: s(T)}
    s_\text{RH(0)}^\text{SM}=\frac{2\pi^2}{45} \,g_{*\text{s, RH(0)}}^\text{SM} \ \left(T_\text{RH(0)}^\text{SM}\right)^3\,,
\end{align}
with $g_{*\text{s, RH}}^\text{SM} =g_\text{SM}=106.75~\mbox{and} ~g_{*s,\,0}^\text{SM}=g_{*s, \,0}=43/11$. Using~\Cref{eq: zeta}, we can then write the total radiation energy density at the end of reheating as
\begin{align}
    \rho_\text{RH}=\frac{\pi^2}{30}\, g_\text{SM}\, \left(T_\text{RH}^\text{SM}\right)^4\left( 1+ \frac{g_\text{DR}}{g_\text{SM}}\zeta^4 \right)\,,\label{eq: TRH}
\end{align}
with $g_\text{DR}$ the effective number of relativistic degrees of freedom of the particles composing the dark radiation abundance at the end of the reheating phase. Then, combining~\Cref{eq: s consv,eq: s(T),eq: TRH}, the third term in \Cref{eq: N} can be rewritten as
\begin{align}\label{eq: 66}
    \ln\left[\frac{a_\text{RH} \,\rho^{1/4}_\text{RH}}{\sqrt{3}\,a_0\,{\cal H}_0}\right] 
    &=\ln\left[\left(\frac{\pi^2}{270} \right)^{1/4}\,\frac{g_{*s, \,0}^{1/3}}{g_\text{SM}^{1/12}}\,\frac{T_0^{\mathrm{SM}}}{{\cal H}_0}\left(1+\frac{g_\text{DR}}{g_\text{SM}}\zeta^4\right)^{1/4}\right]\nonumber\\
    &\simeq 66.50+ \frac{1}{4}\ln\left(1+\frac{g_\text{DR}}{g_\text{SM}}\zeta^4\right) \,,
\end{align}
where we have used $a_0=1$ and $T_0^{\mathrm{SM}}/{\cal H}_0=1.63\times10^{29}$~\cite{Aghanim:2018eyx} in the last equality. Then, inserting \Cref{eq: lnRad1,eq: 66} into \Cref{eq: N}, we obtain a relation between the reheating temperatures of the visible and hidden sectors and the number of e-folds during inflation, namely
\begin{align}
    N_e\simeq 64.48+\frac{1}{3}\ln \frac{T_\text{RH}^{\text{SM}}}{M_\text{Pl}}+\frac{1}{3}\ln \frac{V(\phi_*)}{V(\phi_\text{end})}+\frac{1}{6}\ln\frac{V(\phi_*)}{M^4_\text{Pl}}+\frac{1}{3}\ln\left(1+\frac{g_\text{DR}}{g_\text{SM}}\zeta^4\right)\,,\label{eq: N1}
\end{align}
where we have used $\ln[k_*/{\cal H}_0]=2.182$ for $k_*=0.002\,\text{Mpc}^{-1}$~\cite{Planck:2018jri} and 
\begin{align}\label{eq: rhoend}
    \rho_{\rm end}=\frac{3}{2}V(\phi_\text{end})\quad~\text{and}\quad~{\cal H}_*^4=\frac{1}{9M_\text{Pl}^4}\rho_*^2\simeq \frac{V^2(\phi_*)}{9M_\text{Pl}^4}\,,
\end{align}
with $\varepsilon_\text{end}=1$ and $\varepsilon_* \ll 1$.
%
%
\section{Dark relic abundances}\label{sec: dark relic abundances}
In this section, we study scalaron decays into hidden-sector degrees of freedom, sourced by its coupling to the stress-energy momentum tensor. We obtain general relations for the resulting dark matter and dark radiation relic abundances today and compare them with the measured values.
\subsection{Dark matter}
As argued above, DM candidates in the hidden sector produced from the decays of the inflaton $\phi$ can be studied by neglecting their gravity-suppressed interactions with the SM sector. Their interactions with dark radiation will also be ignored, for simplicity, and will be justified a posteriori. Under these considerations, the coupled system of Boltzmann equations describing the evolution of the inflaton and SM radiation energy densities, denoted $\rho_\phi$ and $\rho_{\rm R}$, respectively, and of the number of DM particles $n_{\rm DM}$, reads~\cite{Chung:1998rq}
\begin{align}
    \frac{\mathrm{d} \rho_\phi}{\rm{dt}} &= -3\,{\cal H}\,\rho_\phi-\Gamma_\phi\, \rho_\phi \,,\label{eq: rhophi}\\
    \frac{\mathrm{d}\rho_\text{R} }{\rm{dt}}&= -4\,{\cal H}\,\rho_\text{R}+ B_\text{R}\,\Gamma_\phi \,\rho_\phi \,,\label{eq: rhoR}\\
    \frac{\mathrm{d}n_\text{DM}}{\rm{dt}}&= -3\,{\cal H}\,n_\text{DM}+ 2\,B_\text{DM}\,\Gamma_\phi 
    \,\frac{\rho_\phi}{m_\phi} \,,\label{eq: rhoDM}
\end{align}
where $B_{\rm DM}$ is the branching ratio of inflaton decays into a pair of DM particles, $B_{\rm R}\,\Gamma_{\phi}$ denotes the energy fraction transferred to SM radiation, and $\Gamma_\phi$ is the total decay width of $\phi$. The Hubble parameter ${\cal H}$ couples the three equations in the above system, since it depends on all energy densities. However, it can also be obtained from the scale factor $a(t)$ through ${\cal H}=\dot{a}/a$, which leads to a solution of~\Cref{eq: rhophi} of the form~\cite{Kolb:1990vq}
\begin{align}\label{eq: Sol-phi}
    \rho_\phi (a(t)) &= \rho_\text{end}\,\left(\frac{a(t_{\mathrm{end}})}{a(t)}\right)^3\, e^{-\Gamma_\phi\,\left(t-t_\text{end}\right)}\,,
\end{align}
where $t_{\mathrm{end}}$ denotes the time at the end of inflation, and $\rho_\text{end}=\rho_\phi(a(t_{\mathrm{end}}))$. To obtain $n_{\rm DM}(a,t)$, we insert the solution in~\Cref{eq: Sol-phi} into~\Cref{eq: rhoDM} and find
\begin{align}\label{eq: rhoDM(a)}
    n_\text{DM}(a,t) = 2\,B_\text{DM}\,\frac{\rho_\text{end}}{m_\phi}\,\left(\frac{a(t_{\mathrm{end}})}{a(t)}\right)^3\,\left(1- e^{-\Gamma_\phi\,(t-t_\text{end})}\right)\,.
\end{align}
The freeze-in value of the dark matter energy density is obtained by evaluating~\Cref{eq: rhoDM(a)} at $t\to\infty$ and multiplying it by the DM mass $m_{\rm DM}$, leading to a relic abundance $\Omega_\text{DM} h^2$ given by
\begin{align}
    \Omega_\text{DM} h^2 &= 2\,B_\text{DM}\, \frac{m_\text{DM}}{m_\phi}\,\left(\frac{a(t_{\mathrm{end}})}{a_0}\right)^3\,\frac{ h^2}{3\,M_\text{Pl}^2\,{\cal H}_0^2}\,\rho_\text{end}\,,
\end{align} 
where the present value of the Hubble parameter is taken to be ${\cal H}_0= h\,2.1332\times 10^{-42}$ GeV, with $h\simeq0.674$~\cite{Aghanim:2018eyx}. 

The ratio $a(t_{\mathrm{end}})/a_0$ can be computed similarly to~\Cref{eq: N1}, which gives
\begin{align}\label{eq: aend}
    \frac{a(t_{\mathrm{end}})}{a_0} &=\frac{a(t_{\mathrm{end}})}{a_\text{RH}}\,\left(\frac{\rho_\text{end}}{\rho_\text{RH}}\right)^{1/4}\left(\frac{a_\text{RH}\,\rho_\text{RH}^{1/4}}{\sqrt{3}\,a_0 \,{\cal H}_0} \right)\left(\frac{\sqrt{3} \,{\cal H}_0}{\rho_\text{end}^{1/4}} \right)\,,
\end{align}
where the product of the first two expressions gives $R_\text{rad}$ as defined in~\Cref{eq: lnR} and the quantity in the third parenthesis is given in~\Cref{eq: 66}. Note that the dependence of $\Omega_\text{DM} h^2$ on $\rho_\text{end}$ cancels because the average equation of state $\bar{w}$ is zero. Finally, we obtain the predicted DM relic abundance today,
\begin{align}
  \Omega_\text{DM} h^2 &=2\sqrt{3} \exp(3\times 66.50)\,B_\text{DM}\,h^2\frac{{\cal H}_0}{M_\text{Pl}}\,\left(\frac{\pi^2}{30}\,g_{\mathrm{SM}}\right)^{1/4}\,\frac{m_\text{DM}}{m_\phi}\,\frac{T_\text{RH}^\text{SM}}{M_{\mathrm{Pl}}}\,\left(1+\frac{g_{\mathrm{DR}}}{g_{\mathrm{SM}}}\zeta^4\right)^{3/4}\nonumber\\
  &\simeq 0.12\, \frac{m_\text{DM}}{m_\phi}\,\left(\frac{B_\text{DM}}{2.9\times 10^{-10}}\right)\,\left(\frac{T_\text{RH}^\text{SM}}{\mathrm{GeV}}\right)\,,\label{eq: Omega}
\end{align}
which is in accordance with the results of Ref.~\cite{Allahverdi:2002nb}.
\subsection{Dark radiation}
In the scenarios of interest, the inflaton will generally not only decay into SM radiation and dark matter, but also into dark radiation, which can contribute significantly to the extra relativistic degrees of freedom $N_\text{eff}$~\cite{Ackerman:2008kmp,Steigman:2012ve,Anchordoqui:2012qu}. Denoting the corresponding partial decay widths $\Gamma_{\rm SM}$, $\Gamma_{\rm DR}$, and $\Gamma_{\rm DM}$, respectively, the associated branching ratios satisfy the relations
\begin{align}
    B_i\equiv \frac{\Gamma_i}{\Gamma_\phi}\,,\qquad
    B_{\rm SM}+B_{\rm DR}+B_{\rm DM}=1\,.
    \label{eq: branching ratios}
\end{align}

Within the perturbative and instantaneous reheating approximation and with the inflaton oscillating around an approximately quadratic minimum, such that it behaves as pressureless matter and the total width remains approximately constant, the fraction of the inflaton energy deposited into a given sector can be obtained by multiplying the total energy density at the end of reheating $\rho_{\mathrm{tot, RH}}$ with the corresponding branching ratio, namely
\begin{align}\label{eq: energy partition branching}
    \rho_i(t_{\rm RH})\simeq B_i\,\rho_{\rm tot,RH}=\frac{\Gamma_i}{\Gamma_\phi}\rho_{\rm tot,RH}\,.
\end{align}
Furthermore, we take the end of reheating to occur when the inflaton decay rate becomes comparable to the Hubble rate. In other words, we identify the total decay width $\Gamma_\phi$ with the Hubble parameter at the end of the reheating phase  ${\cal H}_\text{RH}$~\cite{Kolb:1990vq}, namely
\begin{align}\label{eq: reheating condition}
    \mathcal H_{\rm RH}\simeq\Gamma_\phi\,,
\end{align}
and the residual inflaton energy density is assumed to be negligible at this time. The Friedmann equation then gives
\begin{align}\label{eq: Gamma total}
    \Gamma_\phi^2\simeq{\cal H}_{\rm RH}^2=\frac{\rho_{\rm tot,RH}}{3M_{\rm Pl}^2}\,,
\end{align}
where
\begin{align}\label{eq: total density three sectors}
    \rho_{\rm tot,RH}=\rho_{\rm SM}(T_{\rm RH}^{\rm SM})+\rho_{\rm DR}(T_{\rm RH}^{\rm DR})+\rho_{\rm DM}(t_{\rm RH})\,,
\end{align}
with $\rho_{\rm DM}(t_{\rm RH})$ denoting the full DM energy density at the end of reheating, including the kinetic energy of the particles produced in inflaton decays. Using~\Cref{eq: Gamma total}, this gives~\cite{Chung:1998rq}
\begin{align}\label{eq: Gamma general}
    \Gamma_\phi\Gamma_{\rm SM}=\frac{\rho_{\rm SM}}{3M_{\rm Pl}^2}\,,
    \quad
    \Gamma_\phi\Gamma_{\rm DR}=\frac{\rho_{\rm DR}}{3M_{\rm Pl}^2}\,,
    \quad
    \Gamma_\phi\Gamma_{\rm DM}=
    \frac{\rho_{\rm DM}(t_{\rm RH})}{3M_{\rm Pl}^2}\,.
\end{align}
Notice that these expressions are not independent Friedmann equations. They follow from the total Friedmann equation together with the instantaneous energy-partition assumption in~\Cref{eq: energy partition branching}.

We assume that the SM and DR sectors are internally thermalized, although not necessarily with each other, in which case it is useful to define their ratio as in~\Cref{eq: zeta}. The first two expressions in~\Cref{eq: Gamma general} can be expressed in terms of these temperatures as
\begin{align}
    \Gamma_\phi\Gamma_{\rm SM} &=
    \frac{\pi^2}{90M_{\rm Pl}^2}
    g_{\rm SM}\left(T_{\rm RH}^{\rm SM}\right)^4\,,
    \label{eq: Gamma SM temperature}
    \\
    \Gamma_\phi\Gamma_{\rm DR}&=
    \frac{\pi^2}{90M_{\rm Pl}^2}
    g_{\rm DR}\left(T_{\rm RH}^{\rm DR}\right)^4\,,
    \label{eq: Gamma DR temperature}
\end{align}
which, when combined with the above results, allows us to solve for $\Gamma_{\rm DR}$, namely
\begin{align}\label{eq: GA1}
    \Gamma_{\rm DR}=\left(\frac{g_{\rm DR}^2\pi^2}{90g_{\rm SM}M_{\rm Pl}^2}\right)^{1/2}\left(T_{\rm RH}^{\rm SM}\right)^2\zeta^4\left(1+\frac{g_{\rm DR}}{g_{\rm SM}}\zeta^4+\frac{\Gamma_{\rm DM}}{\Gamma_{\rm SM}}\right)^{-1/2}\,.
\end{align}
Since~\Cref{eq: GA1} yields a relation between $T_\text{RH}^\text{SM}$ and $\zeta$ for a given $\Gamma_\text{DR}$, once the latter is computed and the number of e-foldings is fixed we can combine~\Cref{eq: N1,eq: GA1} to obtain $T_\text{RH}^\text{SM}$ and $\zeta$.

The effect of dark radiation on $N_{\mathrm{eff}}$, parametrized by $\Delta N_{\mathrm{eff}}=N_{\rm eff}-N_{\rm eff}^{\rm SM}$, can be computed based on the temperature of dark radiation at the moment of neutrino decoupling, i.e. at $T=T_\nu$, and by making use of the conservation of entropy per comoving volume. More specifically, from entropy conservation it follows that
\begin{align}
    g_\text{DR}a^3_\text{RH}\left(T^\text{DR}_{\text{RH}}\right)^3 = g_\text{DR}a^3_\nu\left(T^\text{DR}_{\nu}\right)^3\quad \mathrm{and}\quad g_{*s,\, \mathrm{RH}}^{\mathrm{SM}}\,a^3_\text{RH }\left(T^\text{SM}_\text{RH}\right)^3 =g_{*s}(T_\nu)\, a^3_\nu\left(T_\nu\right)^3\,,
\end{align}
with $g_{*s}(T_\nu)=10.75$, which allows us to write the temperature of the hidden sector bath at neutrino decoupling as
\begin{align}
  T_{\nu}^\text{DR} =&\left(\frac{g_{*s}(T_\nu)}{g_{*s,\,\mathrm{RH}}^\text{SM}}\right)^{1/3}\, \zeta\,T_\nu\,.
\end{align}
From this, one can then compute the dark radiation contribution to the total radiation energy density of the Universe at neutrino decoupling, namely
\begin{align}
  \frac{30}{\pi^2}\rho_\nu^\text{DR}
  =g_\text{DR} \,\left(T^\text{DR}_{\nu}\right)^4 = g_\text{DR} \left(\frac{g_{*s}(T_\nu)}{g_\text{SM}}\right)^{4/3}\zeta^4 \,T^4_\nu
  = \Delta N_\text{eff}\left(\frac{7\times 2}{8}\right)\, T_\nu^4\,,
\end{align}
from which it follows that
\begin{align}
  \Delta N_\text{eff}=\left(\frac{4\,g_\text{DR}}{7}\right)\left(\frac{g_{*s}(T_\nu)}{g_\text{SM}}\right)^{4/3}\zeta^4 \simeq 0.027 \,g_\text{DR} \,\zeta^4\,.\label{eq: deltaNeff}
\end{align}
Finally, using the SM prediction $N_{\rm eff}^{\rm SM}=3.044$~\cite{Bennett:2020zkv}, the Planck+BAO~\cite{Aghanim:2018eyx} constraint $N_{\rm eff}=2.99\pm0.17$ implies $\Delta N_{\rm eff}=-0.05\pm0.17$ corresponding to the approximate $1\sigma$ upper bound
\begin{align}\label{eq: Delta N_eff bound}
    \Delta N_{\rm eff}\lesssim0.12\,.
\end{align}
%

\section{Dark matter models}\label{sec: DM models}

In this section, we consider three models for the strongly-interacting hidden sector from which dark matter arises. The corresponding relic abundance as well as the reheating temperature are computed based on the general results derived in~\Cref{sec: dark relic abundances}. Moreover, since the scalaron decays become efficient only at the end of inflation, during the phase of oscillations around the potential minimum, the corresponding fields $\phi$ and $\chi$ in the Einstein and Jordan frames, respectively, can be identified to a good approximation. 

Thus, we can use the simple coupling of $\chi$ to the trace of the energy-momentum tensor $T_{\alpha\beta}$, which can be obtained from the linearized metric coupling to matter
\begin{align}\label{eq: hT}
    \mathcal{L}_{\rm int}=-\frac{1}{2}h_{\mu\nu}T^{\mu\nu}\,,
\end{align}
to calculate the scalaron decay branching ratios needed for determining the generated relic abundances. More specifically, substituting the scalar component in~\Cref{eq: metric expansion 1} one obtains
\begin{align}\label{eq: Lchi}
  {\cal L}_\chi =-\frac{1}{\sqrt{6}}\,\frac{\chi}{M_\text{Pl}}\,T^\alpha_{~\alpha}\,.
\end{align}

Ordinary QCD provides a useful guide for understanding what kinds of particles are expected in a generic confining gauge theory. At low energies, QCD contains pseudo-NGBs (the pions), vector mesons (such as the $\rho$ and $\omega$), baryons, and the anomalously heavy $\eta^\prime$ meson. In QCD, most of these resonances are unstable because lighter hadronic states with identical conserved quantum numbers exist, allowing rapid strong or electroweak decays. Only states protected by exact or accidental symmetries, such as the proton, are stable on cosmological timescales. In contrast, hidden sectors typically possess a much smaller set of interactions with the visible sector. 

As a result, many otherwise allowed decay channels are absent, and hidden-sector states can naturally become stable without introducing additional ad hoc stabilizing symmetries. Therefore, hidden QCD-like theories exhibit analogous spectra to QCD, although the identity of the stable particle depends on the underlying gauge and flavor symmetries. The three models studied below are summarized in~\Cref{tab: DMmodels} and should be viewed as representative realizations of three qualitatively different mechanisms by which stable relics may emerge within the infrared spectra of hidden strong dynamics.
\begin{table}[t]
\centering
\begin{tabular}{p{3.7cm}C{3.15cm}C{3.4cm}C{3.15cm}}
\toprule
     & \textbf{Model I} & \textbf{Model II} & \textbf{Model III}\\
    \midrule Hidden gauge group & $SU(N_c)$ & $SU(N_c)\times SU(2)_L$ & $SU(N_c)\times U(1)_{A^\prime}$\\[1ex]
    
    Particle content & $\psi$ & $\Psi$ & $\Psi$\\[1ex]
    
    Dark matter candidate & $\eta^\prime$ & $Z'$ & $\pi'_\pm$\\[1ex]
    
    Dark radiation & -- & -- & Massless $A^\prime$,\,$\pi^\prime_0$\\[1ex]
    
    Stabilizing mechanism & Lightest composite state; suppressed $\eta^\prime\to g g$ & Gauge and custodial symmetry & Exact conserved $U(1)_{A^\prime}$ charge\\[1ex]
    
    Main observables & $\Omega_{\eta^\prime}$ & $\Omega_{Z'}$ & $\Omega_{\pi'},\, \Delta N_{\rm eff}$\\
\bottomrule
\end{tabular}
\caption{Representative realizations of dark matter in hidden confining sectors. The three models illustrate different symmetry structures and distinct mechanisms responsible for the stability of the dark matter candidate. The fermions $\psi$ and $\Psi$ correspond to a single Dirac fermion and a single Dirac fermion doublet, respectively, and are charged under the corresponding hidden gauge group.}
\label{tab: DMmodels}
\end{table}
\subsection{Model I: Hidden $\eta^\prime$ dark matter}\label{sec: model I}
The first model we consider contains a single fermion $\psi$ belonging to the fundamental representation of the hidden gauge group $SU(N_c)$. The chiral symmetry at the classical level is $U(1)_V\times U(1)_A$, where $U(1)_A$ is assumed to be dynamically broken by strong interactions and gives rise to a single NGB which we denote $\eta^\prime$. Moreover, since $U(1)_A$ is explicitly broken by the chiral anomaly, $\eta^\prime$ acquires a non-vanishing mass $m_{\eta^\prime}$.\footnote{Since $m_{\eta^\prime}$ is not calculable \cite{Witten:1979vv}, we regard it as a free parameter.} Since there are no hidden pions in this scenario, $\eta^\prime$ is expected to be the lightest hidden-sector particle, rendering it stable.\footnote{Decays of $\eta^\prime$ into gravitons are in principle possible, and will be addressed in~\Cref{sec: DM stability I}.}
\subsubsection{Dark matter relic abundance}\label{sec: DM relic abundance I}
Below the confinement scale, the dynamics of NGBs can be understood in terms of a non-linear sigma model, which we describe in~\Cref{ap: non-linear sigma model} of the Appendix. Using the general result in~\Cref{eq: NGB conformal trace} for the contribution of NGBs to the trace of the energy-momentum tensor expanded to quadratic order and evaluated at $\xi=-1/6$, and performing the replacements
\begin{align}
    \pi^a\to\eta^\prime\qquad\mathrm{and}\qquad M_{ab}^2\to m_{\eta^\prime}^2\,,
\end{align}
one obtains
\begin{align}\label{eq: nonlinear eta trace quadratic}
    T^\mu_{\ \mu}=m_{\eta^\prime}^2\eta^{\prime\,2}+\mathcal O\left(\frac{\eta^{\prime\,2}(\partial\eta^\prime)^2}{f_{\eta^\prime}^2},\frac{m_{\eta^\prime}^2\eta^{\prime\,4}}{f_{\eta^\prime}^2}\right)\,,
\end{align}
where $m_{\eta^\prime}$ is a nonperturbatively-induced mass. Then, combining~\Cref{eq: Lchi,eq: nonlinear eta trace quadratic}, the effective interaction relevant for $\chi\rightarrow\eta'\eta'$ is given by
\begin{align}\label{eq: scalaron eta interaction}
    \mathcal L_{\chi\eta'\eta^\prime} =-\frac{1}{\sqrt{6}}\frac{m_{\eta'}^2}{M_{\rm Pl}}\chi(\eta')^2\,,
\end{align}
from which one can compute the corresponding squared matrix element
\begin{align}\label{eq: scalaron eta amplitude}
    \left|\mathcal M(\chi\rightarrow\eta'\eta')\right|^2=
    \frac{2}{3}
    \frac{m_{\eta'}^4}{M_{\rm Pl}^2}\,.
\end{align}
For a scalaron of mass $m_\phi$ decaying into two identical particles of mass $m_{\eta'}$, the two-body decay width is
\begin{align}\label{eq: scalaron eta width}
    \Gamma_{\eta'}=\frac{1}{2m_\phi}\frac{1}{2!}
    \int d\Phi_2\,\left|\mathcal M(\chi\rightarrow\eta'\eta')\right|^2=\Gamma_{\eta'}=\frac{1}{48\pi}\frac{m_{\eta'}^4}{M_{\rm Pl}^2m_\phi}\left(1-\frac{4m_{\eta'}^2}{m_\phi^2}\right)^{1/2}\,,
\end{align}
where the factor $1/2!$ accounts for the identical final-state particles. 

\begin{figure}[t!]
    \centering
        \centering        
        \includegraphics[width=0.5\textwidth]{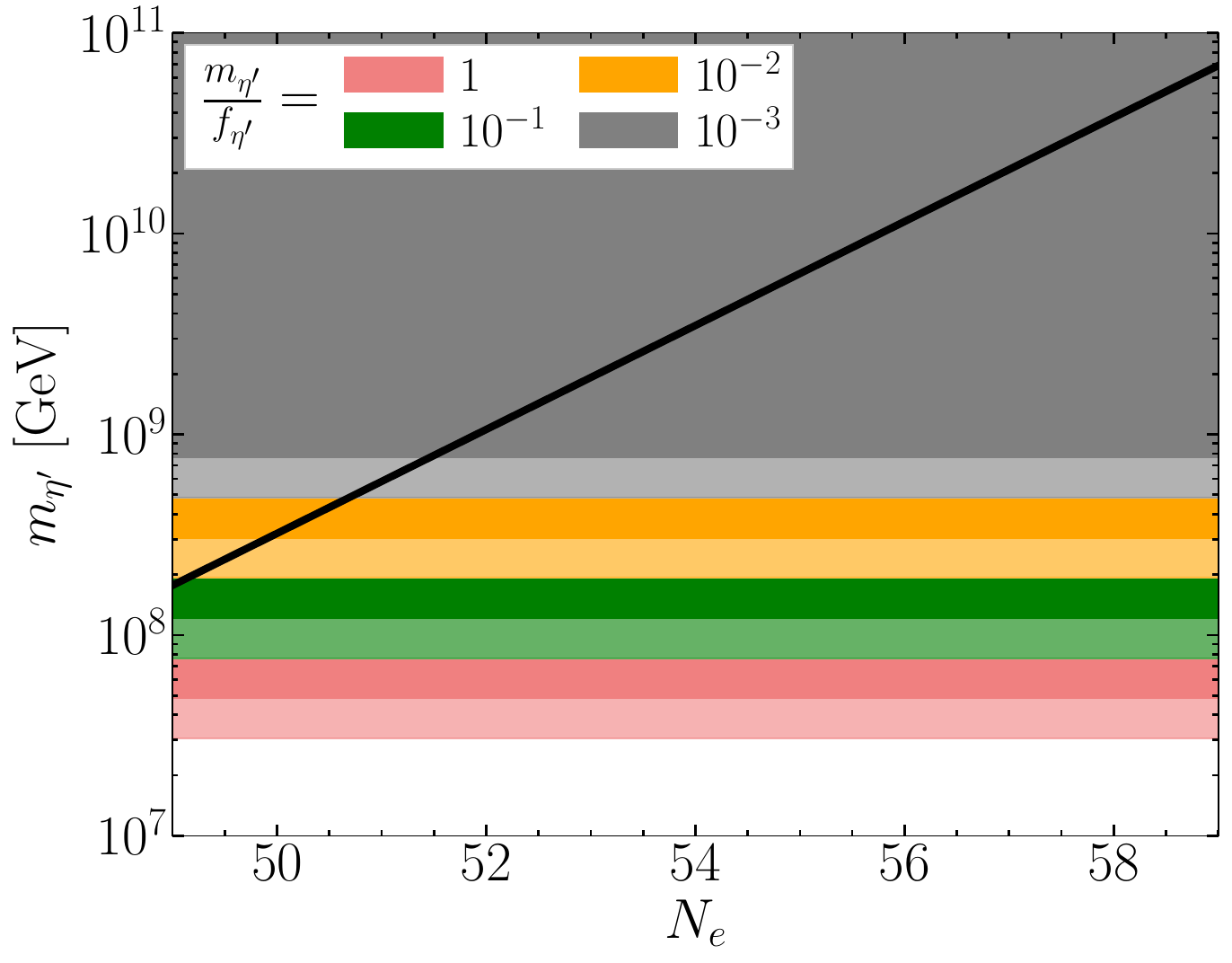}
    \caption{The black line shows the hidden $\eta^\prime$ mass needed to reproduce the observed dark matter relic abundance, as a function of the number of e-folds $N_e$ during inflation. The relic abundance is computed from the effective scalaron-$\eta^\prime$ interaction and the reheating temperature, which can be related to $N_e$ by requiring the Starobinsky inflationary parameters to agree with observations. The dark/light shaded regions are excluded by comparing the estimated lifetime of $\eta^\prime$, which can decay into gravitons, to one time/ten times the age of the Universe. We have set $N_c=3$, and the different colors correspond to different ratios between the mass of $\eta^\prime$ and its decay constant.}
    \label{fig:model1}
\end{figure}

Following Refs.~\cite{Kolb:1990vq,Chung:1998rq}, we identify the inverse of the total decay width of the scalaron with the time scale at which the reheating phase ends (see~\Cref{eq: Gamma total}). Moreover, the reheating temperature $ T_\text{RH}^\text{SM}$ for a given number of e-foldings during inflation $N_e$ can be calculated from~\Cref{eq: N1}. Then, substituting~\Cref{eq: Gamma total} in~\Cref{eq: Omega} with $B_{\eta^\prime}=\Gamma_{\eta^\prime}/\Gamma_\phi$ and $m_\text{DM}\to m_{\eta^\prime}$ allows to find the value of $ m_{\eta^\prime}$, as a function of $N_e$, needed to produce the observed dark matter relic abundance today, namely $\Omega_{\eta^\prime}h^2=0.12$. The resulting values of $m_{\eta^\prime}$ as a function of $N_e$ are shown in~\Cref{fig:model1}.
\subsubsection{Dark matter stability}\label{sec: DM stability I}
Although $\eta^\prime$ is the lightest hidden-sector hadron, it is not necessarily absolutely stable. In curved spacetime, the axial current possesses a mixed gravitational anomaly that allows decays of $\eta^\prime$ into gravitons $g$. Therefore, we estimate the decay width associated with the decay $\eta^\prime\to g+g$ based on the analysis of Ref.~\cite{Araki:2008ek}. Under the axial transformation
\begin{align}\label{eq: axial transf}
    \Psi_a \longrightarrow e^{i\alpha\gamma^5}\Psi_a\,,
\end{align}
the pseudo-NGB transform as
\begin{align}\label{eq: eta prime shift}
    \eta^\prime\longrightarrow\eta^\prime+\alpha f_{\eta^\prime}\,,
\end{align}
and the fermionic path-integral measure is not invariant. For one Dirac fermion in the fundamental representation of $SU(N_c)$, the corresponding anomalous Ward identity can be written as
\begin{align}
    \langle D_\mu J_A^\mu\rangle = \frac{N_c}{192\pi^2}\frac{1}{\sqrt{-g}}\, R\tilde R+\cdots\,,
\end{align}
where $D_\mu$ and $J_A^\mu$ denote the covariant derivative and the axial current, respectively, while
\begin{align}
    R\tilde R\equiv\frac12\epsilon^{\mu\nu\rho\sigma}R_{\mu\nu}{}^{\alpha\beta} R_{\rho\sigma\alpha\beta}\,,
\end{align}
and the ellipsis denotes the hidden gauge contribution.\footnote{The precise overall sign and factors of two depend on the conventions used for the axial current and the dual Riemann tensor.}

The low-energy effective action must reproduce the anomalous variation under an axial transformation of the ultraviolet theory. Adopting the nonlinear parametrization of the pNGBs described in~\Cref{ap: non-linear sigma model}, such that $\eta^\prime$ transforms under $U(1)_A$ as in~\Cref{eq: eta prime shift}, anomaly matching leads to the Wess-Zumino-Witten interaction
\begin{align}\label{eq: eta g g}
    \mathcal L_{\eta^\prime RR}=-\frac{N_c}{192\pi^2}\frac{\eta^\prime}{f_{\eta^\prime}} R\tilde R\,,
\end{align}
which mediates the decay of $\eta^\prime$ into two gravitons, in analogy to the decay of the SM neutral pion into a pair of photons. Expanding the metric according to
\begin{align}\label{eq: metric expansion}
    g_{\mu\nu} = \eta_{\mu\nu}+\frac{2}{M_{\rm Pl}}h_{\mu\nu}\,,
\end{align}
the linearized Riemann tensor scales as
\begin{align}
    R_{\mu\nu\rho\sigma}^{(1)}\sim\frac{\partial^2h}{M_{\rm Pl}}\,.
\end{align}
Consequently, the on-shell decay amplitude is parametrically given by
\begin{align}
    \mathcal M(\eta^\prime\rightarrow gg)\sim\frac{N_c}{92\pi^2}\frac{m_{\eta^\prime}^4}{f_{\eta^\prime}M_{\rm Pl}^2}\,,
\end{align}
and the corresponding width can be estimated as (see~\Cref{app: eta to gg} in the Appendix for more details)
\begin{align}
    \Gamma_{\eta^\prime\rightarrow gg}&\simeq\frac{N_c^2}{\pi(384\pi^2)^2}\frac{m_{\eta^\prime}^7}{f_{\eta^\prime}^2M_{\rm Pl}^4}\nonumber\\
    &\simeq 2.2\times10^{-8}\, N_c^2\,m_{\eta^\prime}\left(\frac{m_{\eta^\prime}}{M_{\rm Pl}}\right)^4\left(\frac{m_{\eta^\prime}}{f_{\eta^\prime}}\right)^2\,.
\end{align}
The shaded regions in~\Cref{fig:model1} exclude values of $m_\eta^\prime$ and $N_e$ for which the lifetime of $\eta^\prime$ obtained from the above estimate is shorter than one or ten times the age of the Universe. We also show how the excluded region reduces as the ratio between the mass of $\eta^\prime$ and its decay constant decreases.
\subsection{Model II: Hidden vector dark matter}\label{sec: Model II}
The second model we consider is based on a hidden sector with a symmetry group given by
\begin{align}\label{eq: symmetry group MII}
  SU(N_c) \times SU(2)_L\times U(1)_L\times SU(2)_R\times SU(1)_R\,,
\end{align}
where only $SU(N_c)$ and $SU(2)_L$ are gauged. We assume the existence of a single hidden-sector fermion doublet
\begin{align}\label{eq: psi_a}
  \Psi_a =\left\{\begin{array}{c}U_a\\D_a\end{array}\right\}\,,
\end{align}
with $a=1,\dots, N_c$ the color index. In the spirit of technicolor models~\cite{Weinberg:1975gm,Susskind:1978ms,Lane:2002wv}, the $SU(N_c)$ gauge theory is vector-like, while the $SU(2)_L$ gauge theory is chiral. Furthermore, to avoid the Witten $SU(2)$ anomaly~\cite{Witten:1982fp}, we assume that $N_c$ is even.

Note that the chiral symmetry $SU(2)_L\times SU(2)_R$ is not explicitly broken by gauging $SU(2)_L$. Moreover, we may assume that the chiral condensate is $SU(2)_V$-invariant, namely
\begin{align}
  \langle \bar{U}_a U_a\rangle=\langle \bar{D}_a D_a\rangle=-N_c\times 4 \pi f_{\pi'}^3\,.
\end{align}
Three massless NGBs arise from the spontaneous breaking of $SU(2)_L\times SU(2)_R$, which are ``eaten'' by the gauging of $SU(2)_L$, and the corresponding gauge bosons acquire a mass (see e.g. Ref.~\cite{Lane:1993wz})
\begin{align}\label{eq: M_Z^prime}
  M_{Z'}= \frac{1}{2}g_{Z^\prime}f_{\pi'}\,.
\end{align}
Thus, the present setup contains no massless scalars, and the mass of the gauge boson $M_{Z'}$ can be much smaller than the Planck mass $M_\text{Pl}$. 
\subsubsection{Dark matter relic abundance}
Unlike the model described in~\Cref{sec: model I}, the mass term of the gauge boson does not contribute to the trace anomaly $T^\alpha_{~\alpha}$, because the mass is generated through spontaneous scale symmetry breaking. The leading effective interaction, needed for calculating the partial decay width $\Gamma_{Z'}$ of $\chi$ into two $Z'$s, is given by
\begin{align}\label{eq: LZ}
  \mathcal{L}_{\chi\,Z^\prime}=-\frac{1}{\sqrt{6}}\,\frac{\chi}{M_\text{Pl} }\,\frac{\beta_{g_{Z^\prime}}}{2g_{Z^\prime}} (Z'_{\alpha\beta})^2\,,
\end{align}
where $Z'_{\alpha\beta}$ is the field strength, and $\beta_{g_{Z^\prime}}$ is the $\beta$-function of the hidden gauge boson coupling. For a gauged $SU(2)_L$ theory with $N_c$ left-handed Weyl doublets the one-loop $\beta$-function for the corresponding gauge coupling reads
\begin{align}\label{eq: beta gZ}
    \beta_{g_{Z'}}=\frac{g_{Z'}^3}{16\pi^2}\left(-\frac{22}{3}+\frac{N_c}{3}\right)\,.
\end{align}
The decay amplitude for $\chi(q)\rightarrow Z'(k_1,\epsilon_1)Z'(k_2,\epsilon_2)$, where $\{q,\,k_1,\,k_2\}$ are the corresponding particle momenta and $\epsilon_{1,2}$ the vector boson polarization vectors, can be obtained from~\Cref{eq: LZ} and is given by
\begin{align}\label{eq: scalar vector amplitude}
    \mathcal M=-4i\frac{1}{\sqrt6 M_{\rm Pl}}\frac{\beta_{g_{Z^\prime}}}{2g_{Z^\prime}}\left[(k_1\cdot k_2)(\epsilon_1^*\cdot\epsilon_2^*)-(k_1\cdot\epsilon_2^*)(k_2\cdot\epsilon_1^*)\right]\,.
\end{align}
After summing over the physical polarizations of the two massive gauge bosons, one obtains
\begin{align}
    \Gamma(\chi\rightarrow Z'Z')=\,&\frac{3}{24\pi}\left(\frac{\beta_{g_{Z'}}}{2g_{Z'}}\right)^2\frac{m_\phi^3}{M_{\rm Pl}^2}\left(1-4x_{Z'}+6x_{Z'}^2\right)\sqrt{1-4x_{Z'}}\nonumber\\
    \simeq\,& \frac{3}{24\pi}\left(\frac{\beta_{g_{Z^\prime}}}{2g_{Z^\prime}}\right)^2\, \frac{m_\phi^3}{M^2_\text{Pl}}\quad\mbox{for}\quad x^2=\frac{M_{Z'}^2}{m_\phi^2}\ll 1\,,
  \label{eq: GZp}
\end{align}
where the factor of three accounts for the three $SU(2)_L$ gauge bosons.

\begin{figure}[t!]
    \centering
        \centering
        \centering
        \includegraphics[width=0.49\textwidth]{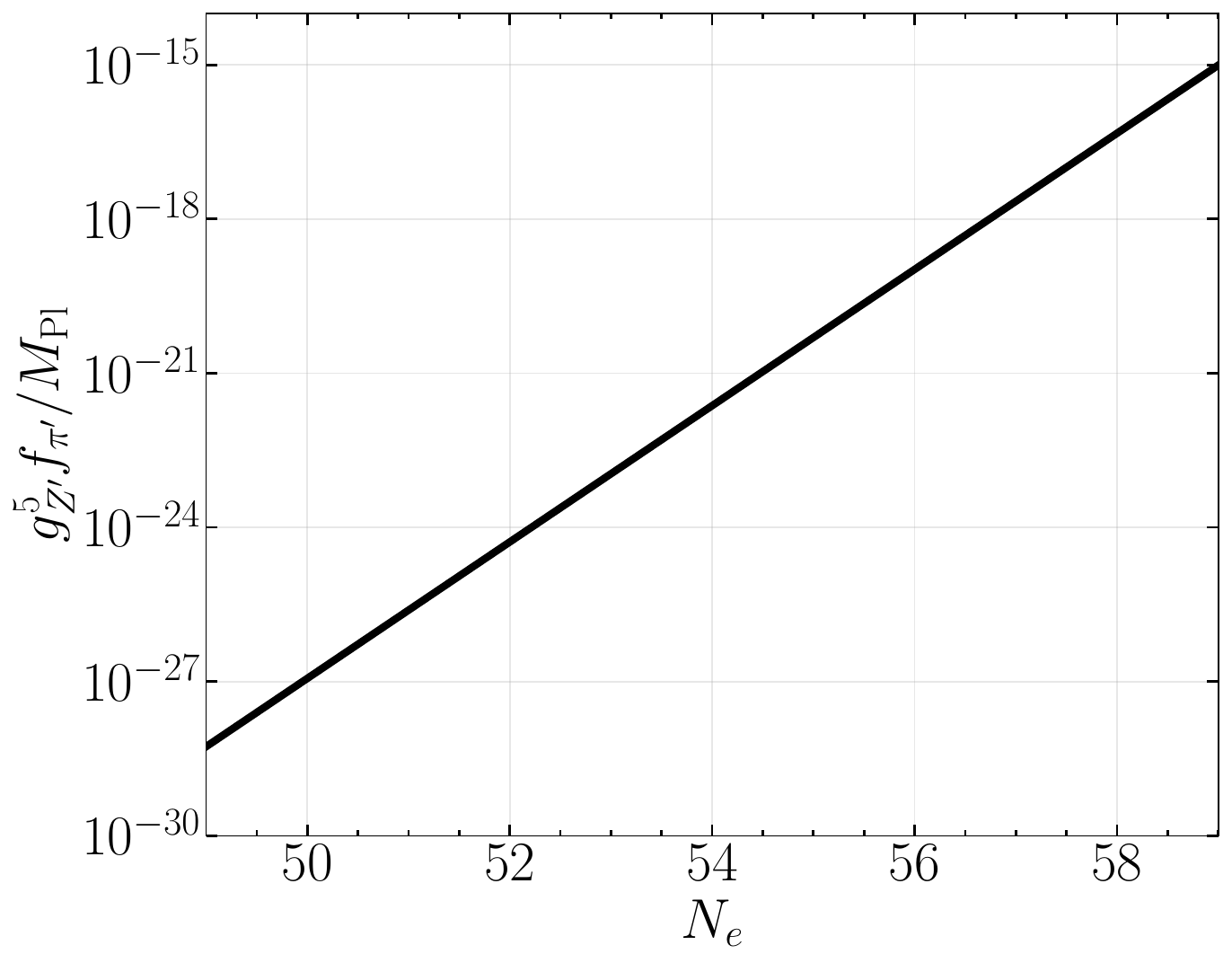}
        \includegraphics[width=0.49\textwidth]{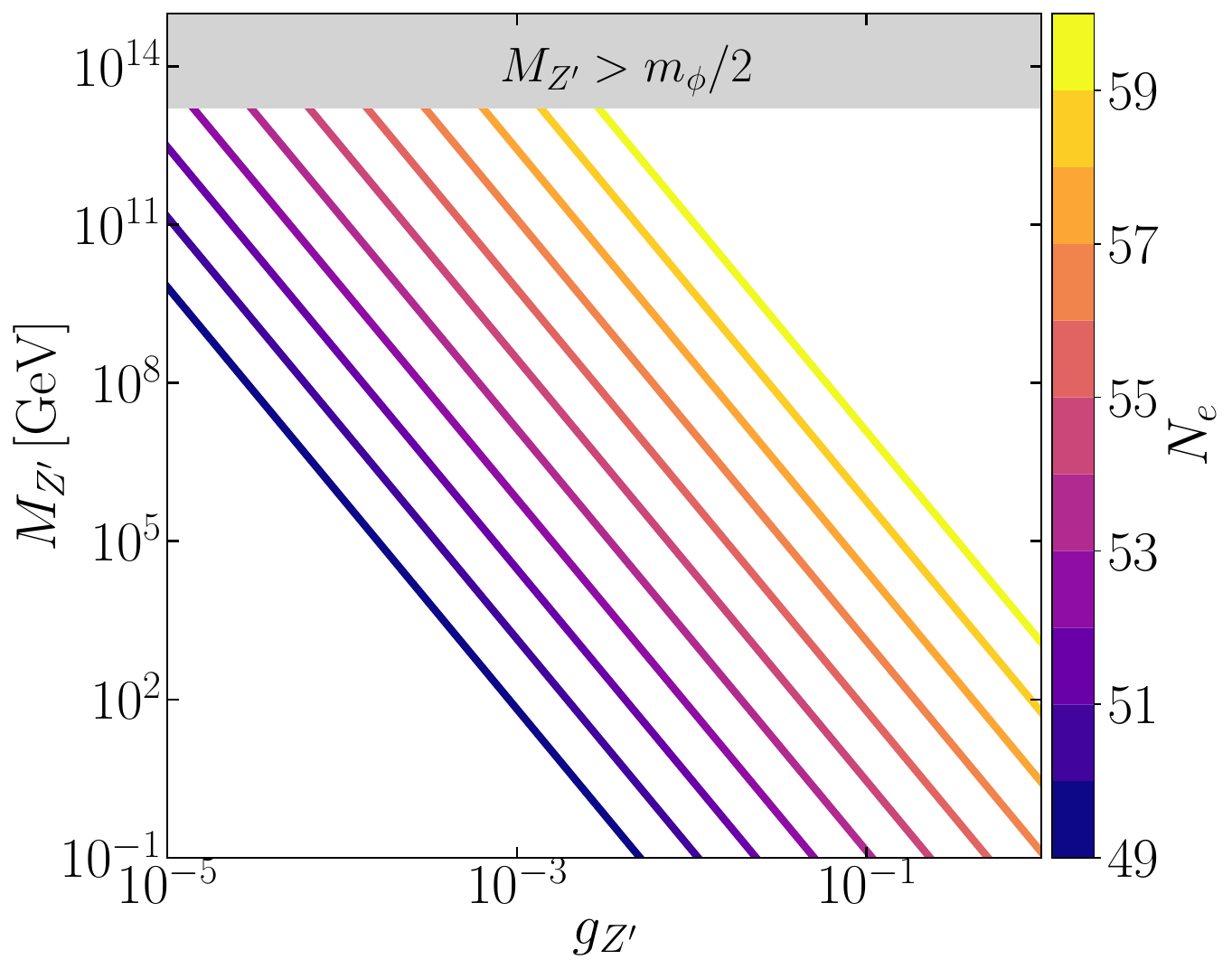}
    \caption{\textit{Left:} Required value of the product $g_{Z'}^5 f_{\pi'}/M_{\rm Pl}$ needed to obtain the correct dark matter abundance today, as a function of the number of e-foldings during inflation $N_e$. This is obtained by fixing $N_c=4$ and assuming $M_{Z'}< m_\phi/2$. \textit{Right:} Values of the hidden-sector $Z^\prime$ boson mass and of the corresponding gauge coupling needed to reproduce the correct dark matter abundance, for different values of $N_e$.}
    \label{fig:model2}
\end{figure}

For definiteness, we consider the case $N_c=4$ and use the one-loop $\beta$-function shown in~\Cref{eq: beta gZ} to evaluate~\Cref{eq: GZp}. This allows to obtain an expression for the dark matter relic abundance via~\Cref{eq: Omega}, as a function of these independent parameters $f_{\pi'}$ and $g_{Z^\prime}$.\footnote{Note that the branching ratio $\tilde{B}_{Z'}$ is $(2\Gamma_{Z'})/\Gamma_\phi$, and the right-hand side of~\Cref{eq: Omega} should be multiplied with $3$ because there are three gauge bosons.} The result of this evaluation is shown in~\Cref{fig:model2}, where we plot the values of the couplings needed to obtain the correct dark matter abundance, as a function of the number of e-foldings during inflation.
\subsubsection{Dark matter stability}
The three massive gauge bosons are stabilized by the unbroken custodial symmetry of the hidden sector. To see this, the chiral condensate may be represented by the matrix
\begin{align}
    \Sigma_{ij}\sim\left\langle\overline{\Psi}_{Rj}\Psi_{Li}\right\rangle\,,
\end{align}
which transforms according to
\begin{align}
    \Sigma \longrightarrow L\Sigma R^\dagger\,,
\end{align}
with $L\in SU(2)_L$ and $R\in SU(2)_R$. The flavor-symmetric condensate assumed above is proportional to the identity,
\begin{align}
    \langle\Sigma\rangle\propto \mathbf 1_{2\times2}\,,
\end{align}
and therefore realizes the symmetry-breaking pattern
\begin{align}
    SU(2)_L\times SU(2)_R\longrightarrow SU(2)_V\,,
\end{align}
where $SU(2)_V$ is the diagonal custodial subgroup. Since $SU(2)_L$
is gauged, the three Nambu--Goldstone bosons are eaten and the corresponding gauge fields $Z_\mu^{\prime A}$, with $A=1,2,3$, acquire the common mass in~\Cref{eq: M_Z^prime}. The massive vectors transform as a triplet of the unbroken custodial symmetry, whereas the graviton, the scalaron, the hidden radial modes and the SM fields are custodial singlets. It then follows that the decay of a single hidden vector into custodial singlets is forbidden.

This conclusion is also manifest from the minimal coupling to gravity shown in~\Cref{eq: hT}. Custodial invariance requires the vector contribution to the energy-momentum tensor to involve contractions such as $\delta^{AB}Z^{\prime A}Z^{\prime B}$. Gravitational interactions therefore contain at least two hidden vectors, generating vertices such as $hZ'Z'$ and $hhZ'Z'$, but not vertices containing one $Z'$ and only gravitons. Pair annihilation or production, $Z'Z'\leftrightarrow gg$, is consequently allowed, whereas the decay $Z'\rightarrow gg$ is forbidden.

Finally, the non-abelian self-interactions do not spoil the stability. Although vertices involving three hidden vectors are present, custodial symmetry enforces the degeneracy of the triplet, so that a decay of one component into two others is kinematically forbidden. The stability is therefore exact provided that no additional interactions distinguish the $U$ and $D$ fermions or otherwise break the custodial $SU(2)_V$ symmetry.
\subsection{Model III: Hidden pseudo-NGB dark matter}\label{sec: model III}
In this third model, we consider the same symmetry group as in~\Cref{sec: Model II} (see~\Cref{eq: symmetry group MII}) and the same particle content, namely a single hidden fermion doublet $\Psi$, defined in~\Cref{eq: psi_a}. However, instead of gauging $SU(2)_L$, we consider gauging a $U(1)_{A^\prime}$ subgroup of $SU(2)_V$. More specifically, we assign a $U(1)_{A^\prime}$ charge of $+1$ to the up component $U_a$ of $\Psi$ and $-1$ to its down component $D_a$. Under this charge assignment, the symmetry-breaking pattern is
\begin{align}
    \Big[U(1)_L\times U(1)_R\Big]\times \Big[U(1)_L\times U(1)_R\Big]\rightarrow U(1)_V\times U(1)_V\,,
\end{align}
and results in three NGBs $\pi'_i$, with $i=1,\,2,\,3$. The $U(1)_{A^\prime}$ interaction shifts the mass of $\pi'_\pm\equiv(\pi'_1\mp i \pi'_2)/\sqrt{2}$ to a non-vanishing value $m_{\pi^\prime_{\pm}}$ due to the explicit breaking of the chiral symmetry by the $U(1)_{A^\prime}$ interaction while $\pi'_0$ remains massless \cite{Dashen:1969eg}. Furthermore, the charged hidden pions $\pi'_\pm$ are stable since the $U(1)_{A^\prime}$ gauge symmetry remains unbroken by the chiral condensate, rendering them viable dark matter candidates. On the other hand, the neutral dark pion $\pi^\prime_0$ and the gauge boson $A_\mu^\prime$ remain exactly massless and, as a consequence, contribute to the dark radiation abundance.
\subsubsection{Dark matter relic abundance}
We begin by considering the production of neutral and charged hidden pions. The contribution of the former to dark radiation is expected to be suppressed in the conformal regime of interest, as found in the analysis of Ref.~\cite{Hill:1991jc} based on the NJL model~\cite{Nambu:1961tp,Nambu:1961fr}.

To estimate the corresponding abundances, we consider the non-linear realization of the NGBs coupled to a background gravity in the conformal regime (see~\Cref{ap: non-linear sigma model} for details). First, we consider the two-body decay of $\chi$ into the $\pi'_\pm$ dark matter. The corresponding masses $m_{\pi'_{\pm}}$ are generated via the $U(1)_{A^\prime}$ gauge interactions and can be calculated within the NJL model (see e.g. Ref.~\cite{Dmitrasinovic:1992hb}), giving
\begin{align}
     m^2 _{\pi'_\pm} = \frac{e_{A^\prime}^2}{4\pi^2}\, \Delta^2\,,
\end{align}
where $\Delta$ is a function of NJL parameters such as the coupling constants and the cut-off scale. However, for the present analysis, we regard $\Delta$ as an independent parameter.

Similar to Model I, the non-vanishing mass terms of $\pi'_\pm$ present in~\Cref{eq: trace5} generate an effective interaction that can be obtained by combining~\Cref{eq: Lchi,eq: NGB conformal trace}, namely
\begin{align}\label{eq: Lpi}
  \mathcal{L}_{\chi\,2\pi^\prime}=-\frac{1}{\sqrt{6}}\,\frac{\chi}{M_\text{Pl} }\, 2 \,m^2_{\pi^\prime_{\pm}}\,\pi'_+ \pi'_-\,,
\end{align}
from which one can calculate the two-body partial decay rate of $\chi\to\pi'_++\pi'_-$, which reads
\begin{align}\label{eq: Gpi}
  \Gamma_{\pi^\prime_\pm}=\frac{1}{24\pi}\frac{m^4_{\pi^\prime_\pm}}{M_\text{Pl}^2m_\phi}\left( 1- 4 \frac{m^2_{\pi^\prime_{\pm}}}{m_\phi^2}  \right)^{1/2}\,.
\end{align}
Next, we consider the decays of $\chi$ into neutral hidden pions. Since they are massless, they do not contribute to the energy-momentum tensor at quadratic order, and to obtain its effective coupling with $\chi$ we need to expand the corresponding effective action in~\Cref{eq: general NGB action} to quartic order. To this end, we expand the functions $G_{ab}(\pi^\prime)$ and $\mathcal F(\pi^\prime)$ defined in~\Cref{eq: G and F} to quadratic order, namely
\begin{align}
  G_{ij}(\pi')=&\delta_{ij}\left(1- \frac{\pi'_k\pi'_k}{3f^2_{\pi'}}\right)+\frac{\pi'_i\pi'_j}{3f^2_{\pi'}}+O(\pi'^4)\,,\\
  \mathcal F(\pi')=&\pi^\prime_j\pi^\prime_j\left(1-\frac{\pi'_i \pi'_i}{12f^2_{\pi'}}\right)+O(\pi'^4)\,.
\end{align}
Plugging this into the general result for the trace of the on-shell energy-momentum tensor given in~\Cref{eq: NGB trace offshell}, we obtain
\begin{align}
  T^{\alpha}_{~\alpha}=&
  -g^{\alpha\beta}\,\partial_\alpha \pi'_i \partial_\beta \pi'_i\,\left(1+6\xi-\left(\frac{1}{3}+5 \xi\right)\frac{\pi'_j\ \pi'_j}{f^2_{\pi'}}\right) -\frac{g^{\alpha\beta}\pi'_i\partial_\alpha \pi'_i \,\pi'_j\partial_\beta \pi'_j}{f^2_{\pi'}}\,\left(\frac{1}{3}+2\xi\right)\nonumber\\
  &+\xi \,R \,\pi'_i \pi'_i\left(1+6\xi-\left(\frac{1}{12}+2\xi\right)\frac{\pi'_j\pi'_j}{f^2_{\pi'}}\right)
  +O(\pi'^6)\,,
\end{align}
which for $\xi=-1/6$ reduces to
\begin{align}\label{eq: trace5}
  T^{\alpha}_{~\alpha}=\frac{1}{2}g^{\alpha\beta}\,\partial_\alpha \pi'_i \partial_\beta \pi'_i\,\frac{\pi'_j \pi'_j}{f^2_{\pi'}}-\frac{1}{24}R \,\pi'_i \pi'_i\,\frac{\pi'_j \pi'_j}{f^2_{\pi'}}+O(\pi'^6)\,.
\end{align}
\begin{figure}[t!]
    \centering
    \includegraphics[width=0.49\textwidth]{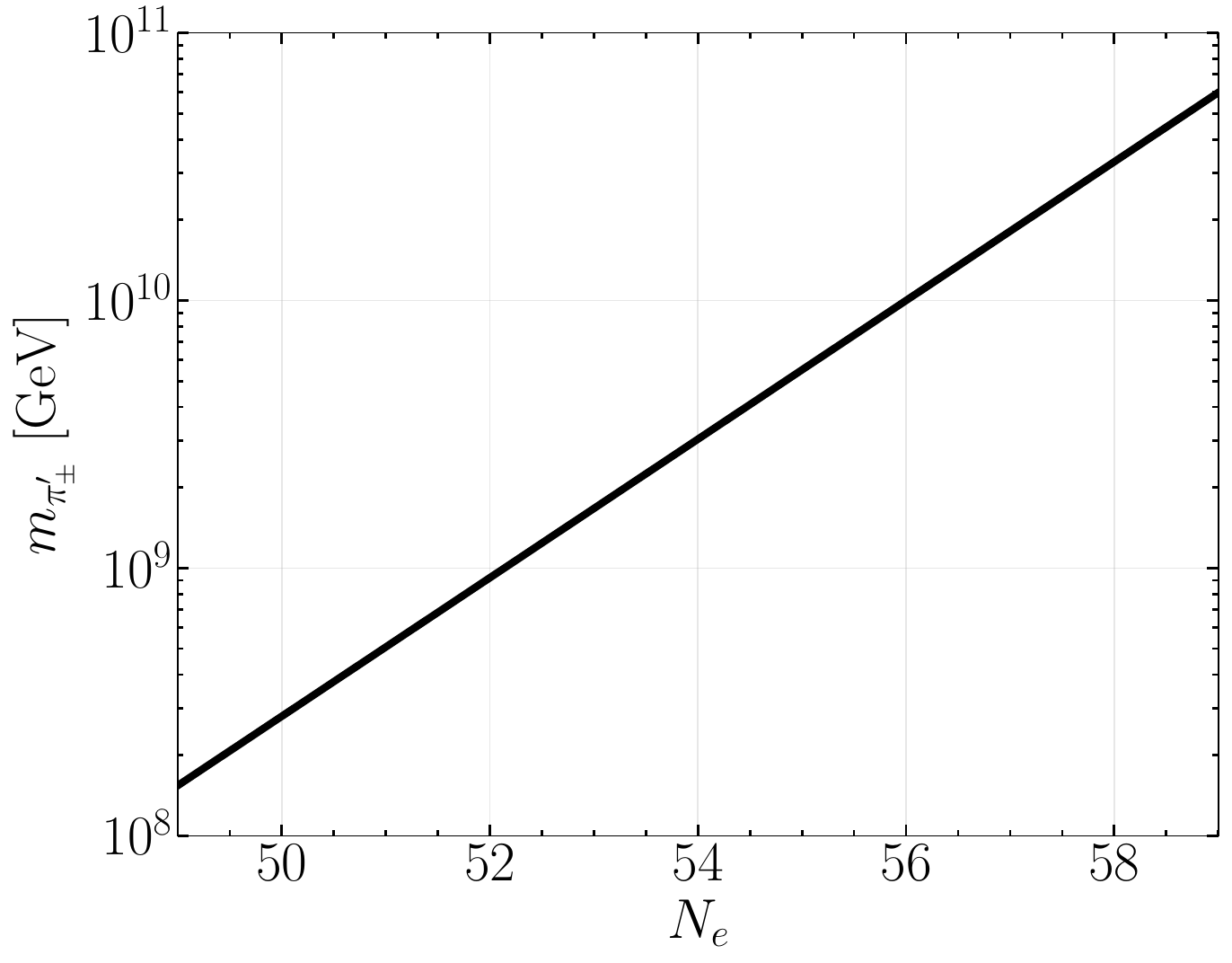}
    \includegraphics[width=0.49\textwidth]{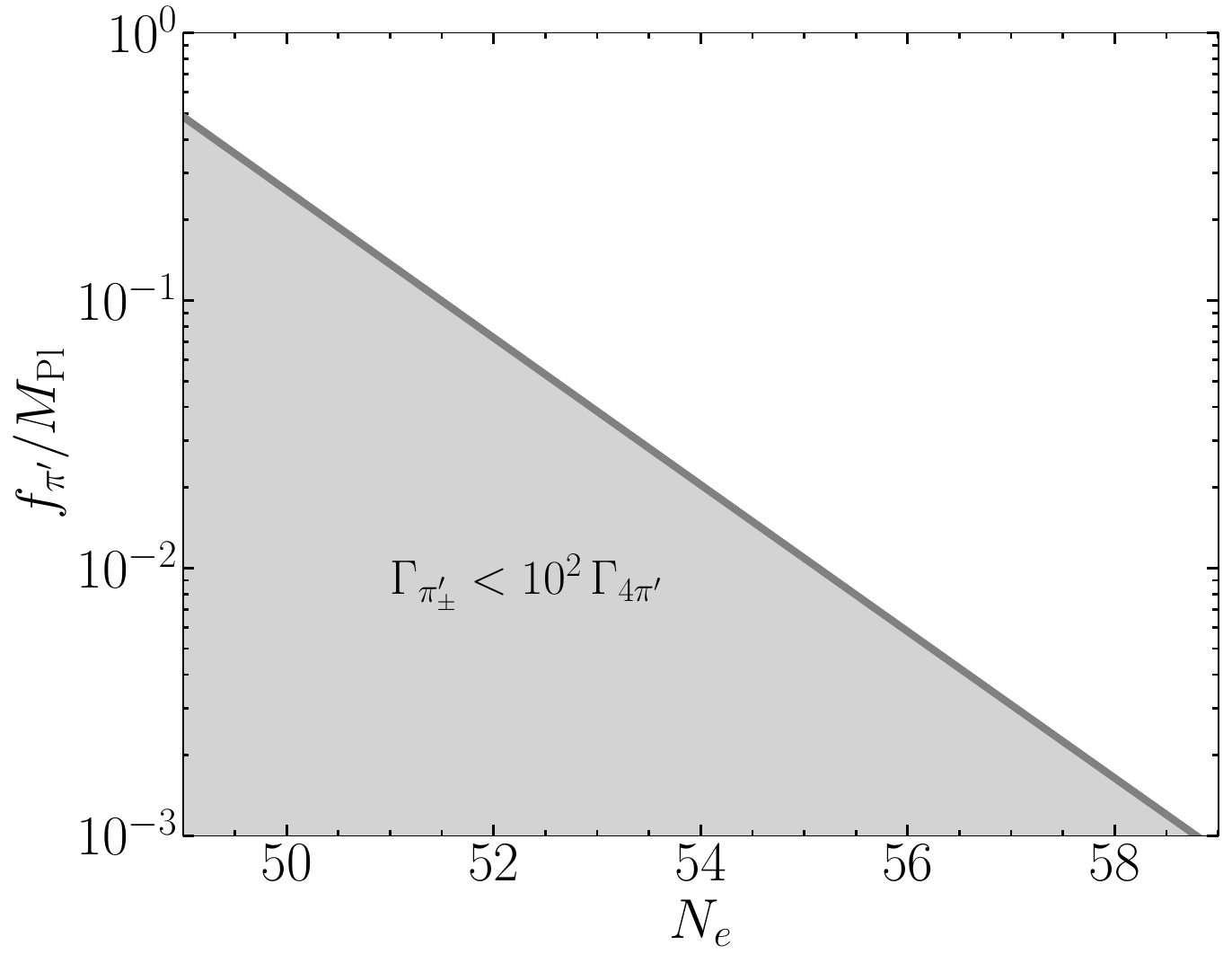}
    \caption{\textit{Left:} Charged hidden pion mass required to reproduce the observed dark matter abundance as a function of the number of e-folds during inflation, neglecting the contribution of 4-body scalaron decays. \textit{Right:} Lower bound on $f_{\pi'}/M_{\rm Pl}$ obtained by requiring $\Gamma_{\pi^\prime_\pm}>10^2\,\Gamma_{4\pi'}$. Below this curve, the four-body scalaron decays into hidden pions should be included.}
    \label{fig:model3}
\end{figure}
Combining~\Cref{eq: trace5,eq: Lchi}, we obtain effective interactions that allow the scalaron $\chi$ to decay into four hidden pions. To estimate the associated decay width, we assume, for simplicity, that all $\pi'$s are massless and that the derivatives acting on $\pi'$ may be replaced by $m_\phi$ (the mass of $\chi$). For a decay process with characteristic momentum of order $m_\phi$, one may estimate
\begin{align}\label{eq: derivative dimensional estimate}
    \partial_\mu\pi_i'\partial^\mu\pi_i'\sim m_\phi^2\pi_i'\pi_i',
\end{align}
up to momentum-dependent and order-one numerical factors, which gives the parametric interaction
\begin{align}\label{eq: L4pi}
  \mathcal{L}_{\chi\,4\pi^\prime}=\frac{1}{M_\text{Pl} }\frac{m_\phi^2}{f^2_{\pi'}}\,\chi (\pi'_i\pi'_i)^2\,.
\end{align}
Four-body decays into massless particles have a large phase space suppression that can be written in closed form as
\begin{align}
    \Phi_4=\frac{m_\phi^4}{2(4\pi)^5\Gamma(4)\Gamma(3)}\sim( 1\times10^{-7})\,m_\phi^4\,,
\end{align}
so that the decay width will be 
\begin{align}\label{eq: G4pi}
  \Gamma_{4\pi'}\sim 10^{-7}\times \frac{m_\phi^3}{M^2_\text{Pl}}\,\left(\frac{m_\phi}{f_{\pi'}}\right)^4\,.
\end{align}
Whether or not these four-body decays are relevant when estimating the final dark matter and dark energy abundances depends on how dominant the two-body decays into $\pi^\prime$s and $A_\mu^\prime$s are. 

In the left panel of~\Cref{fig:model3}, we plot the values of $ m_{\pi^\prime_{\pm}}$ giving the correct dark matter abundance $\Omega_{\pi^\prime_\pm}h^2=0.12$, as a function of the number of e-foldings during inflation. We have assumed the two-body decay width $\Gamma_{\pi^\prime_\pm}$ to be significantly larger than the four-body decay width $\Gamma_{4\pi^\prime}$, so that the latter can be neglected. This assumption is valid for large enough values of the hidden pion decay constant $f_{\pi^\prime}$. The right panel of~\Cref{fig:model3} shows the lower bound on $f_{\pi'}/M_\text{Pl}$ needed to satisfy the condition
\begin{align}
    \Gamma_{\pi^\prime_\pm} > 10^2\times\Gamma_{4\pi'}\,,
\end{align}
and for parameter space points lying below this lower bound, one should also include the four-body decay when calculating $\Omega_{\pi'_\pm}$.
\subsubsection{Dark radiation}
As mentioned above, the hidden-sector gauge boson $A_\mu^\prime$ remains exactly massless and can contribute to the dark radiation abundance. This contribution can be estimated based on the terms in the trace of the energy-momentum tensor proportional to $A^\prime_\mu$ and~\Cref{eq: Lchi}.
\begin{figure}[t!]
    \centering
        \centering
        \includegraphics[width=0.5\textwidth]{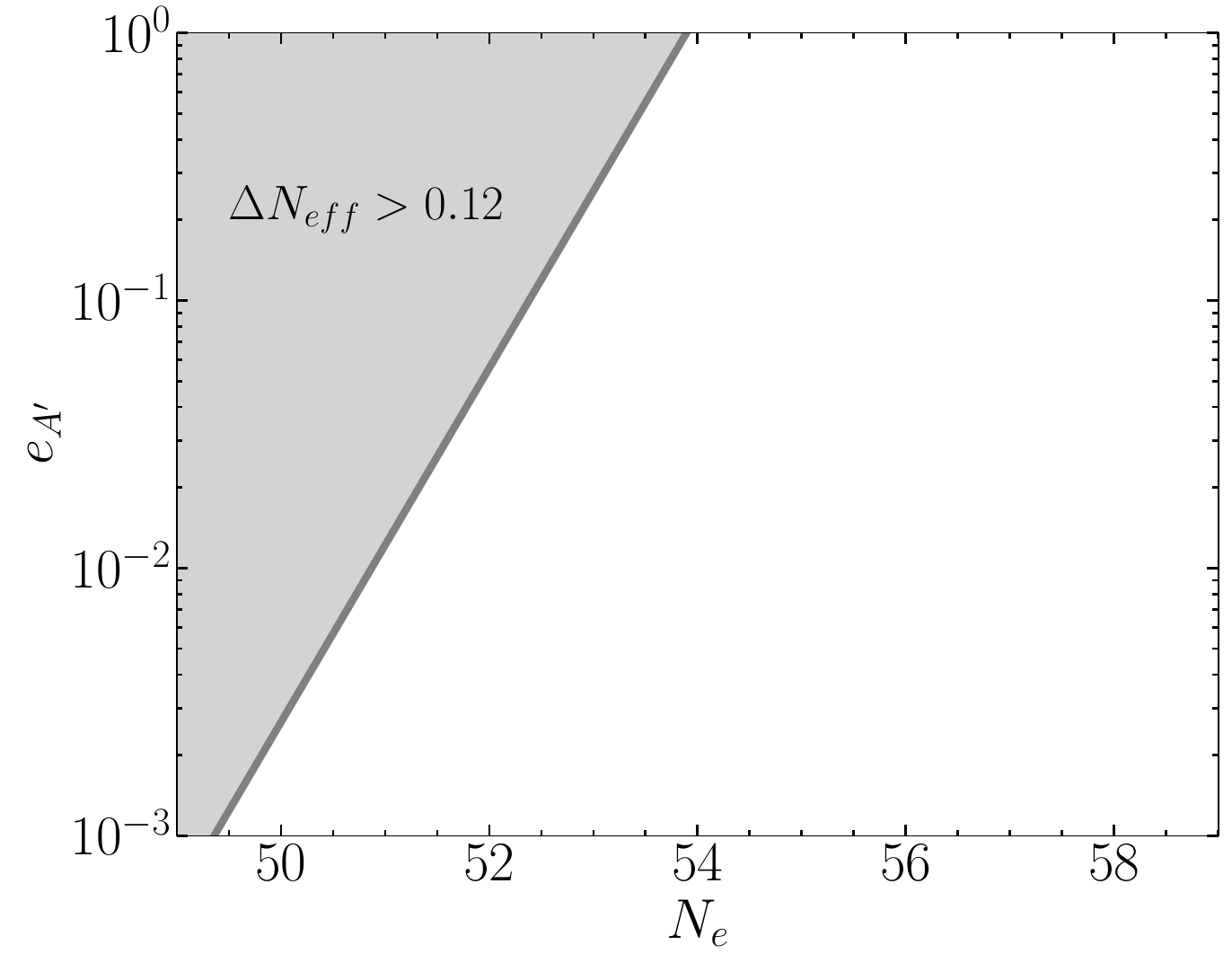}
    \caption{Values of the hidden gauge coupling $e_{A^\prime}$ compatible with the experimental bound on $\Delta N_{\rm eff}$ shown in~\Cref{eq: Delta N_eff bound}, as a function of $N_e$ and with $N_c=4$.}
    \label{fig:max}
\end{figure}

In complete analogy to~\Cref{eq: LZ}, the effective interaction term between $\chi$ and $A_\mu^\prime$ is given by
\begin{align}\label{eq: LA}
  \mathcal{L}_{\chi\,A_\mu^\prime}=-\frac{1}{\sqrt{6}}\,\frac{\chi}{M_\text{Pl} }\, 
  \frac{\beta_{e_{A^\prime}}}{2e_{A^\prime}}(A^\prime_{\alpha\beta})^2\,,
\end{align}
where $\beta_{e_{A^\prime}}$ is the $\beta$-function of the $U(1)_{A^\prime}$ gauge coupling $e_{A^\prime}$, which at one-loop reads
\begin{align}
    \beta_{e_{A^\prime}}^{(1)}=\frac{N_c}{6\pi^2}\,e_{A^\prime}^3\,.
\end{align} 
With~\Cref{eq: LA}, one can compute the partial decay width of the scalaron into a pair of hidden gauge bosons, which is given by
\begin{align}\label{eq: GA2}
  \Gamma_{A^\prime} &=\frac{1}{24\pi}\left(\frac{\beta_{e_{A^\prime}}}{2e_{A^\prime}}\right)^2\frac{m_\phi^3}{M^2_\text{Pl}}\,.
\end{align}
Replacing~\Cref{eq: GA2} in~\Cref{eq: GA1} with $\Gamma_\text{DR}=\Gamma_{A^\prime}$ and $g_\text{DR} =2$, and combining the result with~\Cref{eq: N1} gives an expression for $\zeta \,(= T^{A^\prime}_\text{SM}/T_\text{RH}^\text{SM}) $ in terms of $e_{A^\prime}$, $N_e$, and $N_c$. In~\Cref{fig:max} we show the values of $e_{A^\prime}$ satisfying the experimental bound in~\Cref{eq: Delta N_eff bound}, as a function of $N_e$ and for $N_c=4$.
%
%
\section{Summary and conclusions}\label{sec: Summary and conclusions}
In this work, we have explored the possibility that cosmic inflation, electroweak symmetry breaking, and dark matter originate from a common underlying framework based on classically scale-invariant quadratic gravity coupled to a strongly interacting hidden sector. Classical scale invariance provides an appealing solution to the hierarchy problem~\cite{Bardeen:1995kv} by forbidding fundamental mass parameters at the classical level, such that all physical scales must emerge dynamically~\cite{Coleman:1973jx,Gildener:1976ih}. Quadratic gravity~\cite{Stelle:1976gc,Fradkin:1981iu,Salvio:2018crh} naturally realizes this idea, since it contains only dimensionless couplings and predicts the scalaron associated with the $R^2$ operator, which acts as the inflaton responsible for Starobinsky inflation. The introduction of a hidden confining gauge sector then provides a dynamical origin for the Planck and electroweak scales through dimensional transmutation, while simultaneously giving rise to stable hidden-sector states that naturally constitute dark matter candidates~\cite{Hur:2011sv,Holthausen:2013ota,Kubo:2014ova,Kubo:2018vdw,Hatanaka:2016rek}.
In the conformal limit, the Standard Model Higgs decouples from the inflationary dynamics, while the composite scalar induced by the hidden strong dynamics can be described by an effective Nambu--Jona-Lasinio potential~\cite{Nambu:1961tp,Nambu:1961fr}. After performing the Weyl transformation to the Einstein frame, we showed that the multi-field scalar potential reduces, under well-defined conditions, to an effectively single-field system~\cite{Aoki:2021skm}. The composite scalar remains stabilized near the minimum of its potential during inflation, while the scalaron follows the standard Starobinsky potential~\cite{Starobinsky:1980te}. Consequently, the inflationary predictions are those of Starobinsky inflation, whereas the subsequent reheating phase is driven by scalaron decays into both visible- and hidden-sector degrees of freedom through the universal coupling of the scalaron to the trace of the energy-momentum tensor~\cite{Gorbunov:2010bn}. 

Having constructed visible and hidden sectors that communicate only through gravity, we derived general expressions for the gravitational freeze-in production of dark matter from scalaron decays, together with the corresponding dark-radiation abundance~\cite{Weinberg:2013kea}. The resulting relic abundances depend only on the scalaron branching ratios into hidden particles and the reheating temperature, which can be related to the number of inflationary $e$-folds by taking into account constraints on the inflationary observables. We then made use of these generic results to investigate three representative realizations of the hidden sector. The first model consists of a single-flavor hidden $SU(N_c)$ gauge theory in which the anomalously massive hidden $\eta^\prime$ meson acts as the dark matter candidate. The second model extends the hidden gauge symmetry by a gauged $SU(2)_L$, whose massive vector bosons become the stable dark matter particles. Finally, we considered a hidden sector with an unbroken $U(1)$ gauge symmetry, giving rise simultaneously to charged hidden pions that constitute the dark matter and to massless gauge bosons and neutral pions contributing to dark radiation.

In all three cases, we determined the dark matter mass required to reproduce the observed relic abundance as a function of the reheating history. We found that viable composite dark matter masses within models I and III lie in the range $10^8$--$10^{11}\,\mathrm{GeV}$, depending on the realization of the hidden sector and the duration of inflation, while the corresponding dark-radiation abundance provides an additional cosmological probe of the hidden dynamics. For model I, in particular, we found an upper bound on the mass of $\eta^\prime$, which, despite being the lightest composite state, can decay into gravitons, by requiring its lifetime to be larger than the age of the Universe. On the other hand, hidden gauge bosons in model II are stable due to a remnant custodial symmetry, and the values of their masses needed to constitute the observed dark matter relic abundance span a wide range depending on the values of the corresponding gauge coupling and the duration of inflation. An upper bound of $M_{Z^\prime}\lesssim \mathcal{O}(10^2)\,\mathrm{GeV}$ is found for $g_{Z^\prime}\sim\mathcal{O}(1)$, but which increases quickly to $M_{Z^\prime}\lesssim \mathcal{O}(10^{10})\,\mathrm{GeV}$ for $g_{Z^\prime}\sim\mathcal{O}(10^{-2})$.

Finally, the inflationary observables of the framework studied in this work are those of Starobinsky inflation (up to possible modifications from the Weyl-squared term), while the relic abundances of dark matter and dark radiation are directly linked to the reheating dynamics. Future measurements of the tensor-to-scalar ratio, the effective number of relativistic degrees of freedom, $\Delta N_{\rm eff}$, and improved determinations of inflationary parameters will therefore further constrain the parameter space of the present framework. Although the hidden sector may remain inaccessible to collider experiments, cosmological observations provide a powerful window into the dynamics responsible for generating the fundamental scales of nature. More generally, the framework illustrates how inflation, scale generation, electroweak symmetry breaking and dark matter may all emerge from a single classically scale-invariant theory without introducing ad hoc mass scales or stabilizing symmetries.
%
%
\acknowledgments{
J. P. G. acknowledges funding from the International Max Planck Research School for Precision Tests of Fundamental Symmetries (IMPRS-PTFS). This work was supported in part by the JSPS KAKENHI Grant Number 23K03383 (J.K.).}
%
%
\appendix
\section*{Appendices}
\section{Nonlinear realization of the hidden Nambu--Goldstone sector}\label{ap: non-linear sigma model}
We describe the Nambu--Goldstone bosons associated with the spontaneous breaking of a global symmetry $G\to H$ by defining
\begin{align}\label{eq: general nonlinear field}
    U(\pi)=\exp\left(\frac{2i\pi}{f_\pi}\right)\qquad\mathrm{and}\qquad\pi=\pi^a T^a\,,
\end{align}
where $T^a$ are the broken generators and $f_\pi$ denotes the corresponding decay constant. We consider the following nonlinear effective action in the presence of a gravitational background,
\begin{align}\label{eq: general NGB action}
    S_{\rm NGB}=\int d^4x\sqrt{-g}\left[\frac{f_\pi^2}{4\mathcal N}g^{\mu\nu}{\rm Tr}\left(\partial_\mu U^\dagger\partial_\nu U\right)+\frac{\xi f_\pi^2}{4\mathcal N}R\,{\rm Tr}\left(U^\dagger+U-2\mathbb I\right)-V_{\rm br}(U)\right]\,,
\end{align}
where $\mathcal N$ is a normalization factor determined by the normalization of the generators, while $V_{\rm br}$ parametrizes interactions that explicitly break the global symmetry (for an exact NGB, $V_{\rm br}=0$). This action can then be expressed in field-space form by expanding the nonlinear action in terms of the coordinates $\pi^a$, namely
\begin{align}
    S_{\rm NGB}=\int d^4x\sqrt{-g}\left[\frac12 g^{\mu\nu}G_{ab}(\pi)\partial_\mu\pi^a\partial_\nu\pi^b-\frac{\xi}{2}R\,\mathcal F(\pi)-V_{\rm br}(\pi)\right]\,,
    \label{eq: general-NGB-field-space-action}
\end{align}
where the field-space metric $G_{ab}$ and the curvature function $\mathcal{F}$ are defined by
\begin{align}\label{eq: G and F}
    \frac12G_{ab}(\pi)\partial_\mu\pi^a\partial^\mu\pi^b&\equiv\frac{f_\pi^2}{4\mathcal N}{\rm Tr}\left(\partial_\mu U^\dagger\partial^\mu U\right)\,,\\
    \mathcal F(\pi)&\equiv-\frac{f_\pi^2}{2\mathcal N}{\rm Tr}\left(U^\dagger+U-2\mathbb I\right)\,.
\end{align}
Near the vacuum $\pi^a=0$, the normalization may be chosen such that
\begin{align}\label{eq: general small field expansion}
    G_{ab}(0)=\delta_{ab}\,,\qquad\mathcal F(\pi)=\pi^a\pi^a+\mathcal O\left(\frac{\pi^4}{f_\pi^2}\right)\,.
\end{align}
The energy-momentum tensor can be obtained by taking the variation of the action with respect to the metric, namely
\begin{align}\label{eq: Tmunu}
    T_{\mu\nu} =\frac{2}{\sqrt{-g}}\frac{\delta S_{\rm NGB}}{\delta g^{\mu\nu}}\,.
\end{align}
For the action in~\Cref{eq: general-NGB-field-space-action} it evaluates to
\begin{align}
    T_{\mu\nu}=\,&\,G_{ab}(\pi)\partial_\mu\pi^a\partial_\nu\pi^b-g_{\mu\nu}\left(\frac12 G_{ab}(\pi)\partial_\rho\pi^a\partial^\rho\pi^b-V_{\rm br}(\pi)\right)\nonumber\\
    &-\xi\left(\mathcal{G}_{\mu\nu}\mathcal F(\pi)+\left(g_{\mu\nu}\Box-\nabla_\mu\nabla_\nu\right)\mathcal F(\pi)\right)\,,
    \label{eq: general NGB EMT}
\end{align}
where 
\begin{align}
    \mathcal G_{\mu\nu}=R_{\mu\nu}-\frac12g_{\mu\nu}R
\end{align}
is the Einstein tensor. Then, taking the trace in four spacetime dimensions gives
\begin{align}
    T^\mu_{\ \mu}=-G_{ab}(\pi)\partial_\mu\pi^a\partial^\mu\pi^b+4V_{\rm br}(\pi)+\xi\left(R\mathcal F(\pi)-3\Box\mathcal F(\pi)\right)\,,
    \label{eq: NGB trace offshell}
\end{align}
which can be simplified by using the equations of motion for the nonlinear fields, which are
\begin{align}\label{eq: NGB eom}
    \Box\pi^a+\Gamma^a_{\ bc}(\pi)\partial_\mu\pi^b\partial^\mu\pi^c+G^{ab}(\pi)\frac{\partial V_{\rm br}}{\partial\pi^b}+\frac{\xi}{2}R\, G^{ab}(\pi)\frac{\partial\mathcal F}{\partial\pi^b}=0\,,
\end{align}
where
\begin{align}
    \Gamma^a_{\ bc}=\frac12G^{ad}\left(\partial_bG_{dc}+\partial_cG_{db}-\partial_dG_{bc}\right)
\end{align}
is the Christoffel connection associated with the field-space metric.

For processes involving two Nambu--Goldstone bosons, it is sufficient
to expand the nonlinear theory to quadratic order around the vacuum as
\begin{align}
    G_{ab}(\pi)&=\delta_{ab}+\mathcal O\left(\frac{\pi^2}{f_\pi^2}\right)\,,\\
    \mathcal F(\pi)&=\pi^a\pi^a+\mathcal O\left(\frac{\pi^4}{f_\pi^2}\right)\,,\\
    V_{\rm br}(\pi)&=\frac12M_{ab}^2\pi^a\pi^b+\mathcal O(\pi^3)\,.
\end{align}
The quadratic action relevant for two-body decays is therefore
\begin{align}\label{eq: NGB quadratic action}
    S_{\rm NGB}^{(2)}=\int d^4x\sqrt{-g}\left[\frac12\partial_\mu\pi^a\partial^\mu\pi^a-\frac12M_{ab}^2\pi^a\pi^b+\frac{\xi}{2}R\,\pi^a\pi^a\right]\,,
\end{align}
which after evaluating in~\Cref{eq: Tmunu} and setting the curvature to zero gives
\begin{align}\label{eq: NGB quadratic trace}
    T^\mu_{\ \mu}=-\partial_\mu\pi^a\partial^\mu\pi^a+2M_{ab}^2\pi^a\pi^b-3\xi\Box(\pi^a\pi^a)\,.
\end{align}
Using
\begin{align}
    \Box(\pi^a\pi^a)=2\partial_\mu\pi^a\partial^\mu\pi^a+2\pi^a\Box\pi^a\qquad\mathrm{and}\qquad\Box\pi^a+M_{ab}^2\pi^b=0\,,
\end{align}
leads to the following trace, at quadratic order, of the on-shell and flat-space energy-momentum tensor
\begin{align}\label{eq: NGB onshell trace}
    T^\mu_{\ \mu}=-(1+6\xi)\partial_\mu\pi^a\partial^\mu\pi^a+2(1-3\xi)M_{ab}^2\pi^a\pi^b\,.
\end{align}
For the conformally improved value $\xi=-1/6$, this reduces to
\begin{align}\label{eq: NGB conformal trace}
    T^\mu_{\ \mu}=M_{ab}^2\pi^a\pi^b+\mathcal{O}((\pi^\prime)^4)\,.
\end{align}
%
%
\section{$\eta^\prime$ decays into gravitons}\label{app: eta to gg}
In the following, we compute the leading-order contribution to the decay width of $\eta^\prime$ into a pair of gravitons. The relevant interaction is given in~\Cref{eq: eta g g}, and we expand the metric as shown in~\Cref{eq: metric expansion}. Moreover, the leading-order contribution to the Riemann tensor reads
\begin{align}\label{eq: Riemann}
    R_{\lambda\mu\nu\rho}=\frac{1}{M_{\rm Pl}}\left(\partial_\nu\partial_\mu h_{\lambda\rho}+\partial_\rho\partial_\lambda h_{\mu\nu}-\partial_\rho\partial_\mu h_{\lambda\nu}-\partial_\nu\partial_\lambda h_{\mu\rho} \right)\,.
\end{align}
Next, we assign momenta $k_1$ and $k_2$ to the outgoing gravitons and work in the rest frame of the decaying $\eta^\prime$ particle, such that
\begin{align}
    k_1^\mu=\frac{m_{\eta^\prime}}{2}\left(1,\,0,\,0,\,1\right)\qquad\mathrm{and}\qquad k_2^\mu=\frac{m_{\eta^\prime}}{2}\left(1,\,0,\,0,\,-1\right)\,.
\end{align}
A convenient choice of helicity-one polarization vectors is
\begin{align}\label{eq: vector polarizations}
    \varepsilon_\pm^\mu(k_1)=\frac{1}{\sqrt{2}}(0,1,\pm i,0)\qquad\mathrm{and}\qquad\varepsilon_\pm^\mu(k_2)=\frac{1}{\sqrt{2}}(0,1,\mp i,0)\,,
\end{align}
from which one can build the helicity-two graviton polarization tensors
\begin{align}\label{eq: graviton helicity tensors}
    e_{\mu\nu}^{(\pm2)}(k)=\varepsilon_\mu^\pm(k)\varepsilon_\nu^\pm(k)\,,
\end{align}
that satisfy the relations
\begin{align}\label{eq: e relations}
    k^\mu e_{\mu\nu}^{(\lambda)}(k)=0\,,\qquad
    \eta^{\mu\nu}e_{\mu\nu}^{(\lambda)}(k)&=0\,,\qquad e_{\mu\nu}^{(\lambda)}e^{(\lambda')\,\mu\nu *}=\delta_{\lambda\lambda'}\,.
\end{align}
Accounting for the factor of 2 due to the different ways of assigning the final momenta, the amplitude extracted from~\Cref{eq: eta g g}, within the leading-order expansion in~\Cref{eq: Riemann}, reads
\begin{align}
    -i\mathcal{M}_{\lambda\lambda^\prime}=\frac{iN_c}{192\pi^2f_{\eta^\prime}}\frac{2\epsilon^{\mu\nu\rho\sigma}}{2M_{\mathrm{Pl}}^2}&\big(k_1^\alpha k_{1\nu}e_\mu^{(\lambda)\beta*}(k_1)+k_1^\beta k_{1\mu}e_\nu^{(\lambda)\alpha *}(k_1)\nonumber\\&\quad-k_1^\beta k_{1\nu}e_\mu^{(\lambda)\alpha*}(k_1)-k_1^\alpha k_{1\mu}e_\nu^{(\lambda)\beta *}(k_1)\big)\times\nonumber\\
    &\big(k_{2\alpha} k_{2\sigma}e_{\rho\beta}^{(\lambda^\prime)*}(k_2)+k_{2\beta}k_{2\rho}e_{\sigma\alpha}^{(\lambda^\prime)*}(k_2)\nonumber\\
    &\quad-k_{2\beta}k_{2\sigma}e_{\rho\alpha}^{(\lambda^\prime)*}(k_2)-k_{2\alpha} k_{2\rho}e_{\sigma\beta}^{(\lambda^\prime)*}(k_2)\big)\,.
\end{align}
Using~\Cref{eq: e relations} and
\begin{align}
    \epsilon^{\mu\nu\rho\sigma}k_{1\nu}k_{2\sigma}\varepsilon_\mu^{(s)*}(k_1)\varepsilon_\mu^{(s)*}(k_1)=-is(k_1\cdot k_2)=-is\frac{m_{\eta^\prime}^2}{2}\,,
\end{align}
with $s=\pm$, we obtain ($\lambda=\pm 2$)
\begin{align}
    -i\mathcal{M}_{\lambda\lambda^\prime}=2\frac{\lambda}{2}\frac{m_{\eta^\prime}^4}{M_{\rm Pl}^2}\delta_{\lambda\lambda^\prime}\,.
\end{align}
Finally, we evaluate the decay width of $\eta^\prime$ into two identical massless gravitons, which gives
\begin{align}
    \Gamma_{\eta^\prime}=\frac{1}{2m_{\eta^\prime}}\frac{1}{2!}\int d\Phi_2\sum_{\lambda,\,\lambda^\prime}|\mathcal{M}_{\lambda\lambda^\prime}|^2=\frac{N_c^2\,m_{\eta^\prime}^7}{\pi(384\pi^2)^2f_{\eta^\prime}^2 M_{\rm Pl}^4}\,.
\end{align}
%
%
\bibliographystyle{JHEP}
\bibliography{biblio.bib}
%
%
\end{document}